\documentclass[letterpaper, 10 pt, final]{IEEEtran}

\IEEEoverridecommandlockouts

\usepackage[tbtags,cmex10]{amsmath}
\usepackage{amssymb,amsfonts,amsthm}
\usepackage{eucal,bm}
\usepackage{textcomp,gensymb,subfigure}
\usepackage{booktabs}
\usepackage{times,url}
\usepackage{comment}
\usepackage{color}
\usepackage{amsmath,amssymb,amsthm}
\usepackage[utf8]{inputenc}
\usepackage[T1]{fontenc}
\usepackage{xcolor}
\usepackage[colorlinks=true,pagebackref=false]{hyperref}
\hypersetup{urlcolor=blue,citecolor=blue,linkcolor=blue}
\usepackage[capitalise,noabbrev,nameinlink]{cleveref}
\usepackage{mathtools}

\usepackage{cite}

\usepackage{array}

\usepackage{graphicx}
\usepackage{epstopdf}

\usepackage{enumitem}
\newcommand{\f}[2]{\frac{#1}{#2}}

\theoremstyle{definition}
\newtheorem{theorem}{Theorem}

\newtheorem{assumption}{Assumption}
\newtheorem{remark}{Remark}
\newtheorem{proposition}{Proposition}
\allowdisplaybreaks

\title{Computing Modes of Instability of Parameterized Nonlinear Systems for Vulnerability Assessment}

\author{Jinghan Wang and Michael W. Fisher}

\begin{document}


\maketitle

\begin{abstract}
Engineered systems naturally experience large disturbances which have the potential to disrupt desired operation because the system may fail to recover to a desired stable equilibrium point. It is valuable to determine the mechanism of instability when the system is subject to a particular finite-time disturbance, because this information can be used to improve vulnerability detection, and to design controllers to reduce vulnerability. For a large class of nonlinear systems there exists a particular unstable equilibrium point on the region of attraction boundary of the desired stable equilibrium point such that the unstable eigenvector of the Jacobian at this unstable equilibrium point represents the mode of instability for the disturbance. Unfortunately, it is challenging to find this mode of instability, especially in high dimensional systems, because it is often computationally intractable to obtain this particular unstable equilibrium point for a given disturbance. This paper develops a novel algorithm for numerically computing the mode of instability for parameter-dependent nonlinear systems without prior knowledge of the particular unstable equilibrium point, resulting in a computationally efficient method. The key idea is to first consider the setting where the system recovers, and to average the Jacobian along the system trajectory from the post-disturbance state up until the Jacobian becomes stable. As the system approaches inability to recover, the averaged Jacobians converge to the Jacobian at the particular unstable equilibrium point, and can be used to extract the unstable eigenvector representing the mode of instability. Convergence guarantees are provided for computing the mode of instability, both for the theoretical setting in continuous time, and for the proposed algorithm which relies on numerical integration. Numerical examples illustrate the successful application of the method to identify the mechanism of instability in power systems subject to temporary short circuits.
\end{abstract}

\section{Introduction}\label{sec:intro}
Disturbances naturally occur in engineered systems, and may disrupt
desired system behavior. For example, temporary short circuits in a
power system can lead to blackout conditions. When the system is
unable to recover from a given finite-time disturbance to a stable
equilibrium point (SEP) representing desired behavior, often there is
a particular subset of system states whose coupled dynamics is
responsible for this failure to recover. We refer to these lower order
dynamics that result in failure to recover as the mechanism of
instability of the disturbance \cite{kundur2004definition}. In
networked or complex systems, the mechanism of instability typically
also reveals the components of the system most responsible for this
inability to recover. Determining the mechanism of instability is
valuable because focusing on the most important subset of system
dynamics for a given disturbance facilitates detecting proximity to
vulnerability to the disturbance 
and
designing controllers to reduce this vulnerability
\cite{chiang1993predicting}, \cite{fouad1991power}.





Fix a particular finite-time disturbance, such as a particular short circuit at a particular location in the system, and denote the system state immediately after the disturbance as the post-disturbance initial condition, or initial condition (IC) for short. Whether or not the system recovers from the disturbance to the desired SEP depends on whether the IC lies within the region of attraction of that SEP. When the system is just marginally unable to recover, its IC lies on the region of attraction boundary of the SEP and, for a large class of nonlinear systems, within the stable manifold of some unstable equilibrium point, which is known as the controlling unstable equilibrium point (CUEP) \cite{chiang1988stability}. Generically, the Jacobian at the CUEP has a unique unstable eigenvalue \cite{chiang1993predicting} and, as in \cite{fouad1991power}, we define the nonlinear mode of instability to be the corresponding eigenvector of that unstable eigenvalue. In the situations where the system fails to recover, this mode of instability is responsible for driving the system state away from the region of attraction after the disturbance \cite{fouad1991power}, and thus represents the mechanism of instability.

There is a long history on identifying the nonlinear mode of instability. In \cite{michel1985mechanism} the mode of instability is computed and then used to identify the coupled oscillators in the system that are the first to lose synchronization in response to a disturbance.
More recent work includes using the mode of instability to quantify the relative contributions of each state to transient dynamics using CUEP-based participation factor analysis with its corresponding unstable eigenvector \cite{ma2023dominant}.
The main method in the literature for computing the mode of instability is to first determine the CUEP, and then directly find the unstable eigenvector of its Jacobian \cite{fouad1991power}. However, determining the CUEP is itself a challenging problem and can often be intractable in practice \cite{behera1985analytical}. Although several methods exist in the literature for finding the CUEP, including the mode of disturbance method \cite{fouad1984critical}, the exit point method (also known as the BCU method) \cite{chiang1988foundations}, and the shadowing method \cite{treinen1996improved}, these methods typically lack rigorous convergence guarantees, may converge to an incorrect equilibrium point or fail to converge at all, and often require many time-consuming simulations, which can lead to prohibitively large computational cost, especially for high dimensional systems.

The main goal of this paper is to develop a computationally efficient method with rigorous convergence guarantees for computing the nonlinear mode of instability without requiring knowledge of the CUEP. This provides guarantees of reliability for correctly computing the mode of instability while avoiding the challenge of identifying the CUEP. While much of the historical work focused on finding the mode of instability for a single vector field, this paper considers the more general and more challenging case of finding this mode for a parameter-dependent vector field with parameter-dependent IC. This more general setting is adopted here because disturbance recovery often depends on system parameters, which may be uncertain and time-varying. As in \cite{fisher2023stability}, we define the recovery boundary to be the boundary in parameter space between the parameter values for which the system recovers from the disturbance and the parameter values for which it fails to recover. In the special case where the parameter is chosen to be the IC, the recovery boundary becomes equal to the region of attraction boundary, so it can be interpreted as a generalization of the region of attraction boundary to assess vulnerability in parameter space.

The method presented in this paper exploits the properties that as parameter values approach the recovery boundary in parameter space, the IC approaches the region of attraction boundary in state space and the amount of time the system trajectory spends near the CUEP diverges towards infinity \cite{fisher2022hausdorff}. For a parameter value for which the system recovers from the disturbance, the key idea of the proposed approach is to average the Jacobian along the system trajectory from the IC until the final time at which the Jacobian transitions from unstable to stable. The intuition is that as parameter values approach the recovery boundary, since the system trajectory spends increasing time near the CUEP, the Jacobians along the system trajectory will spend increasing time close to the Jacobian at the CUEP, and their average will approach the Jacobian at the CUEP. However, it is important to only average along the trajectory until the final time at which the Jacobian transitions to becoming stable, since the remaining portion of the trajectory converges to the SEP and, thus, averaging over the entire trajectory would result in the average approaching the Jacobian at the SEP rather than the Jacobian at the CUEP.

The method presented in this paper relies on prior work \cite{fisher2019numerical}, \cite{fisher2025computing} to find parameter values on the recovery boundary, and numerically computes this average of the Jacobians as parameter values approach the recovery boundary. We prove that as the parameter values approach the recovery boundary, this average of the Jacobians converges to the true Jacobian at the CUEP. Furthermore, under the same conditions we prove that the average of the Jacobians will have a unique unstable eigenvalue whose corresponding eigenvector converges to the mode of instability. The average of the Jacobians is evaluated in practice using numerical integration of the underlying system trajectory, which introduces approximation errors. Therefore, for the proposed algorithm which relies on numerical integration, we further prove that the approximation of the average of the Jacobians obtained from numerical integration, and its eigenvector corresponding to its unstable eigenvalue, will converge to the true Jacobian at the CUEP and the mode of instability, respectively, in the limit as the time step of the numerical integration approaches zero. Ultimately, the proposed approach is computationally efficient, does not require knowledge of the CUEP, and these convergence guarantees ensure it will reliably compute the mode of instability accurately.

The method is first validated on the simple example of a nonlinear pendulum subject to a disturbance.  In this case, since the system is low dimensional the CUEP and mode of instability can be easily obtained analytically. It is shown that the proposed method accurately computes the mode of instability.  Then, the method is applied to determine the mode of instability of a power system containing multiple generators subject to a temporary short circuit at a particular location in the network without attempting to identify the CUEP. The points on the recovery boundary are considered both in one and higher dimensional parameter spaces. The identified modes of instability provide non-intuitive insights into the mechanism of instability, including a case where the generator most responsible for failure to recover is distant from the location of the disturbance, a situation where one generator plays a minimal role for the inability to recover, and that the PSS controllers, which are often associated with the failure to recover, do not exhibit much impact in these cases. This unexpected dynamic behavior is challenging to predict or identify without applying the proposed algorithm.

The remainder of the paper is organized as follows. Section \labelcref{sec:not} provides background information. Section \labelcref{sec:res} discusses the main results. Section \labelcref{sec:ex} shows illustrative examples. Section \labelcref{sec:proof} provides the proofs. Finally, Section \labelcref{sec:con} offers concluding remarks.

\section{Background}\label{sec:not}


Let $J$ be a connected smooth manifold that is a
subset of $\mathbb{R}^{m}$ for some integer $m > 0$ representing the parameter space, where
$\mathbb{R}^m$ is the $m$-dimensional Euclidean space.
Consider a family of
vector fields $f:\mathbb{R}^{n}\times J\rightarrow\mathbb{R}^{n}$,
parameterized by $J$, such that
$f$ is $C^{1}$ and $n>0$ is an integer. Then for any $p\in J$, we
have that $f_{p}:=f\left(\cdot,p\right)$ is a vector field on
$\mathbb{R}^{n}$.
We use $\dot{x}$ to denote the time derivative of the dynamic states $x\in\mathbb{R}^n$. We model a
particular finite-time parameter-dependent
disturbance as a
parameter-dependent post-disturbance initial condition, which we
simply refer to as the IC $x_{0}:J\rightarrow\mathbb{R}^{n}$ given by
$x\left(0\right):=x_{0}\left(p\right)$. Hence, we consider the family
of ordinary differential equations (ODEs) given for $p\in J$ by
\begin{align}
\dot{x}&=f\left(x,p\right)\label{eq:1},\\
x\left(0\right)&=x_{0}\left(p\right)\label{eq:2}.
\end{align}
Let $\phi\left(x,p,t\right)$ denote the flow of the
vector field $f$ from the IC $x\in\mathbb{R}^n$ at
time $t\in\mathbb{R}$ for parameter value $p\in J$. 
For any subset $T \subset \mathbb{R}$, let $\phi\left(x,p,T\right) = \bigcup_{t \in T} \phi\left(x,p,t\right)$. For any subsets $J' \subset J$ and $T \subset \mathbb{R}$, let $\phi\left(x_0\left(J'\right),J',T\right)$ denote the disjoint union $\sqcup_{p \in J'} \phi\left(x_0\left(p\right),p,T\right) = \cup_{p \in J'} \phi\left(x_0\left(p\right),p,T\right) \times \left\{p\right\} \subset \mathbb{R}^n \times J$.

Consider a fixed $\tilde{p}\in J$. For any $x\in\mathbb{R}^{n}$ we say
that $x$ is nonwandering under $f_{\tilde{p}}$ if for every open
neighborhood $U$ of $x$ and any $T\in\mathbb{R}$, there exists
$t\in\mathbb{R}$ of the same sign as $T$ with $\left|t\right|>T$ such
that $\phi\left(U,\tilde{p},t\right)\cap U\neq\emptyset$. Let
$\Omega\left(f_{\tilde{p}}\right)$ denote the set of all nonwandering
points of $f_{\tilde{p}}$, and note that this includes all equilibria
and periodic orbits of $f_{\tilde{p}}$, as well as chaotic dynamics
and other forms of recurrent behavior. Let a \textit{critical element}
$x\left(\tilde{p}\right)$ be either an equilibrium point or a periodic
orbit of $f_{\tilde{p}}$. An equilibrium point
$x\left(\tilde{p}\right)$ of $f_{\tilde{p}}$ is hyperbolic if its
linearization $\frac{\partial f}{\partial
  x}\left(x\left(\tilde{p}\right)\right)$ has no purely imaginary
eigenvalues. A periodic orbit $x\left(\tilde{p}\right)$ of
$f_{\tilde{p}}$ is hyperbolic if there exists $x\in
x\left(\tilde{p}\right)$ and a cross section $S$ containing $x$ such
that the Poincare first return map $\tau: S\rightarrow S$ is
well-defined and its linearization $\frac{d\tau\left(x\right)}{dx}$
has no eigenvalues of norm one. Every hyperbolic critical element
$x\left(\tilde{p}\right)$ of $f_{\tilde{p}}$ possesses a stable
manifold $W^{s}\left(x\left(\tilde{p}\right)\right)$ and an unstable
manifold $W^{u}\left(x\left(\tilde{p}\right)\right)$ where
$W^{s}\left(x\left(\tilde{p}\right)\right)$
$\left(W^{u}\left(x\left(\tilde{p}\right)\right)\right)$ consists of
all initial conditions that converge to $x\left(\tilde{p}\right)$ in
forwards (backwards) time, respectively. Furthermore, there exist
local stable and unstable manifolds, denoted by
$W^{s}_{\text{loc}}\left(x\left(\tilde{p}\right)\right)$ and
$W^{u}_{\text{loc}}\left(x\left(\tilde{p}\right)\right)$, which are
invariant in forwards and backwards time, respectively. Furthermore,
for any hyperbolic critical element $x\left(\tilde{p}\right)$ of
$f_{\tilde{p}}$, for $J$ sufficiently small and any $p\in J$, there
exists a unique hyperbolic critical element $x\left(p\right)$ that is
equal to $x\left(\tilde{p}\right)$ at $p=\tilde{p}$, and such that
$x\left(p\right)$, $W^{s}_{\text{loc}}\left(x\left(p\right)\right)$,
and $W^{u}_{\text{loc}}\left(x\left(p\right)\right)$ all vary $C^{1}$
with $p$. For any $J'\subset J$, define
$x\left(J'\right)= \sqcup_{p \in J'} x(p) = \cup_{p\in
  J'}x\left(p\right)\times\left\{p\right\}\subset\mathbb{R}^{n}\times
J$. Define $W^{s}\left(x\left(J'\right)\right)$ and
$W^{u}\left(x\left(J'\right)\right)$ analogously.

The notion of a generic $C^{1}$ vector field is used to describe typical behavior, similar to the concept of probability one in a probability space. A property that holds for a generic class of $C^{1}$ vector fields is therefore considered to represent typical or expected behavior. Since there exists many pathological $C^{1}$ vector fields, it is often beneficial to restrict analysis to certain classes of generic $C^{1}$ vector fields, which facilitates establishing the theoretical framework for generic vector fields that often would not hold for arbitrary vector fields. For any set $A$ in a topological space, let $\partial A$ denote its topological boundary, $\overline{A}$ its topological closure, and $\text{int }A$ its topological interior. For any point $x\in\mathbb{R}^{n}$ in state space and any $r>0$, define $B_{r}\left(x\right)$ to be the open ball of radius $r$ centered at $x$. For any manifold $A$ and $x\in A$, let $T_{x}A$ denote the tangent space to $A$ at $x$. Two manifolds $A,B\subset\mathbb{R}^{n}$ are transverse if for every $x\in A\cap B$, $T_{x}A+T_{x}B=T_{x}\mathbb{R}^{n}$. For additional background in differential topology, including the strong and weak $C^{1}$ topologies, we refer the reader to \cite[Chapter 2]{hirsch2012differential}. For more information regarding $C^{1}$ distance and $\epsilon$ $C^{1}$-close, we refer the reader to the background section in \cite{fisher2022hausdorff}.

Suppose $p_{0}\in J$ such that $f_{p_{0}}$ possesses a hyperbolic stable equilibrium point (SEP), denoted by $x^{s}\left(p_{0}\right)$. Then for $J$ sufficiently small and any $p\in J$, there exists a unique hyperbolic SEP $x^{s}\left(p\right)$ near $x^{s}\left(p_{0}\right)$. For $p\in J$,
we define the region of attraction of $x^{s}\left(p\right)$ to be its
stable manifold $W^{s}\left(x^{s}\left(p\right)\right)$,
and $\partial W^{s}\left(x^{s}\left(p\right)\right)$ denotes the region of attraction boundary. For a parameter value $p\in J$, the system recovers from
the disturbance if and only if
$x_{0}\left(p\right)\in
W^{s}\left(x^{s}\left(p\right)\right)$. 
Let the \textit{recovery region}
$R$ be the set of parameter values in $J$ for which the system
recovers, and define the \textit{recovery boundary} to be its topological
boundary $\partial R$ in $J$. We define parameter values that lie on the recovery boundary as \textit{boundary parameter values}.




Propositions~\ref{pro:1} and \ref{pro:2} will be useful for establishing
rigorous guarantees for our method to compute the nonlinear mode of instability
for a large class of parameterized nonlinear systems.
Proposition \labelcref{pro:1} was proved in \cite{fisher2022hausdorff}
and \cite{fisher2023stability}, and states that under general
assumptions satisfied by a large class of dynamical systems, and for
sufficiently small $J$, the region of attraction
boundary in state space varies continuously with respect to parameter,
and equals the union of the stable manifolds of the critical elements
it contains. Since each parameter value on the recovery boundary has corresponding IC lying on the region of attraction boundary in state space, and therefore in the stable manifold of some critical element,
which we call the \textit{controlling critical element}, and whose unstable
manifold intersects the region of attraction. This proposition establishes a direct connection between the recovery boundary in parameter space and the region of attraction boundary in state space, which will be used for proving the theoretical guarantees of the proposed algorithm for computing the mode of instability for parameterized nonlinear systems.

\begin{proposition}
\label{pro:1}
\textit{\cite[Theorem 4.17, Corollary 4.18, Theorem 4.21, Corollary 4.23]{fisher2023stability} Assume there exists $p_{0}\in J$
  such that:}
\begin{enumerate}[label=(\roman*)]
 \item \textit{Every critical element in $\partial
  W^{s}\left(x^{s}\left(p_{0}\right)\right)$ is hyperbolic.}
 \item \textit{The intersections of the stable and unstable
      manifolds of the critical elements in $\partial
      W^{s}\left(x^{s}\left(p_{0}\right)\right)$ are transverse.}
 \item \textit{The intersection of $\partial
   W^{s}\left(x^{s}\left(p_{0}\right)\right)$ with
   $\Omega\left(f_{p_{0}}\right)$ consists of a finite union of
   critical elements $\left\{x^{i}\left(p_{0}\right)\right\}_{i\in
     I}$, where the set $I$ indexes the critical elements that lie on
   $\partial W^{s}\left(x^{s}\left(p_{0}\right)\right)$.}
 \item
   \textit{There exists a neighborhood of infinity which contains no
     nonwandering points of $f_{p_{0}}$ and no orbits which diverge to
     infinity in both forwards and backwards time.}
 \item
   \textit{The vector field family $\left\{{f_{p}}\right\}_{p\in J}$
     is strong $C^{1}$ continuous and $x_{0}$ is $C^{1}$.}
 \item
   \textit{Certain lower semicontinuous functions over $J$ are
     continuous at $p_{0}$, see \cite[Theorem 4.17]{fisher2023stability} for more details.}
\end{enumerate}
\textit{Then for $J$ sufficiently small, $\partial
  W^{s}\left(x^{s}\left(p\right)\right)$ varies continuously with $p$
  and $\partial W^{s}\left(x^{s}\left(J\right)\right)=\bigcup_{i\in
    I}W^{s}\left(x^{i}\left(J\right)\right)$. For any
  $p^{*}\in\partial R$, $x_{0}\left(p^{*}\right)\in\partial
  W^{s}\left(x^{s}\left(p^{*}\right)\right)$. Then
  $x_{0}\left(p^{*}\right)\in
  W^{s}\left(x^{j}\left(p^{*}\right)\right)$ for a unique $j\in
  I$, $x^{j}\left(p^{*}\right)$ is called the controlling critical element for $p^{*}$, and  $W^{s}\left(x^{s}\left(p^{*}\right)\right)\cap  W^{u}\left(x^{j}\left(p^{*}\right)\right)\neq\emptyset$.}
\end{proposition}

\begin{remark}
Note that Assumptions (i), (ii), and (vi) are generic, and it is generically true that $\Omega(f_{p_0})$ is equal to the closure of the union of the critical elements of $f_{p_0}$ \cite{pugh1983c1} (compare to Assumption (iii)), 
so (iii) is equivalent to a finite number of critical elements, along with this generic condition. Thus, the conditions of Proposition \labelcref{pro:1} are very general and hold for a large class of realistic engineering system models.
\end{remark}

\begin{assumption}
\label{ass:1}
\textit{By Proposition \labelcref{pro:1}, $\partial
  W^{s}\left(x^{s}\left(J\right)\right)=\bigcup_{i\in
    I}W^{s}\left(x^{i}\left(J\right)\right)$. Assume that
  $x_{0}\left(J\right)$ is transverse to
  $W^{s}\left(x^{i}\left(J\right)\right)$ for all $i\in I$
  \cite[Assumption 1]{fisher2025computing}.}
\end{assumption}
\begin{remark}
\label{rem:1}
\textit{Assumption \labelcref{ass:1} is generic because $C^{1}$
  submanifolds are generically transverse
  \cite[Theorem A.3.20]{katok1999introduction}.}
\end{remark}
\begin{proposition}
\label{pro:2}
\textit{Assume the conditions of Proposition \labelcref{pro:1} and
  Assumption \labelcref{ass:1}. Then there exists an open neighborhood
  $N$ of $\partial R$ such that for generic $p_{0}\in N$, there exists
  a unique boundary parameter value $p^{*}\in\partial R$ that is closest to
  $p_0$, and that
  $p^{*}$ varies $C^{1}$ continuously with initial parameter value $p_{0}$
  \cite[Theorem 4]{fisher2025computing}.}
\end{proposition}
\begin{remark}
\label{rem:2}
\textit{Proposition \labelcref{pro:2} shows that for generic initial parameter value $p_{0}$ sufficiently close to the recovery boundary, $p_{0}$ has a unique closest point on the recovery boundary that varies smoothly with respect to $p_{0}$.}
\end{remark}
Recent algorithms have been developed for efficiently computing the closest boundary parameter value $p^* \in \partial R$ to an initial value $p_0 \in R$ in high dimensional parameter and state space \cite{fisher2025computing}.
Throughout this paper, for $p^* \in \partial R$, let $\lim_{p\rightarrow p^{*}}$ denote the limit as $p$ approaches $p^*$ from within $R$, i.e., $\lim_{p \rightarrow p^*,~p \in R}$.
We let $\left\|\cdot\right\|$ denote the Frobenius norm defined as $\left\|A\right\|_{\text{F}}=\sqrt{\sum_{i=1}^{n}{\sum_{j=1}^{n}\left|A_{ij}\right|^{2}}}$.

\section{Main Results}\label{sec:res}

\subsection{Defining Mode of Instability for Parameterized Systems}

We begin by generalizing the definition of the nonlinear mode of instability of a disturbance to parameter-dependent vector fields.
Let $N$ be the neighborhood of $\partial R$ from Proposition~\ref{pro:2}, and fix any initial parameter value $p_{0}\in N$ belonging to the generic set
described in Proposition~\ref{pro:2}. Then there exists a unique closest boundary parameter value $p^* \in \partial R$ to $p_0$.
By Proposition~\ref{pro:1}, there exists a critical element $x^*\left(p^*\right)$ in the region of attraction boundary of $x^s\left(p^*\right)$ such that $x_0\left(p^*\right)$ lies in the stable manifold of $x^*\left(p^*\right)$.

\begin{assumption}
  \label{ass:2}
  \textit{$W^s\left(x^*\left(p^*\right)\right)$ has codimension one.}
\end{assumption}

\begin{remark}
\label{rem:a}
\textit{Assumption~\ref{ass:2} is generic \cite[Lemma 3]{fisher2025computing}.}
\end{remark}

\begin{assumption}
  \label{ass:3}
  \textit{$x^*\left(p^*\right)$ is an equilibrium point.}
\end{assumption}

\begin{remark}
\label{rem:3}
\textit{For many practical nonlinear systems including power systems,
  Assumption \labelcref{ass:3} is ubiquitous and has been observed
  through numerical experiments on a wide range of realistic models \cite{treinen1996improved}.}
\end{remark}

Under Assumptions~\ref{ass:2}-\ref{ass:3}, $x^*\left(p^*\right)$ is known as the CUEP,
and we denote it by $x^u\left(p^*\right)$.
Define $A\left(p^{*}\right)$ to be the Jacobian at $x^{u}\left(p^{*}\right)$:
\begin{equation}
\label{eq:3}
A\left(p^{*}\right):=\frac{\partial f}{\partial x}\left(x^{u}\left(p^{*}\right)\right).
\end{equation}\par
By Assumption~\ref{ass:2} and Proposition~\ref{pro:1}(i), $A\left(p^*\right)$ has a
unique unstable eigenvalue
which we denote by $\lambda\left(p^{*}\right)$.
The {\it mode of instability} is defined to be the eigenvector of $A\left(p^*\right)$
associated with the eigenvalue $\lambda\left(p^{*}\right)$, and is denoted
$v\left(p^{*}\right)$.

\subsection{Obtaining Mode of Instability via Jacobian Averaging}\label{sec:ave}

In order to avoid the complexity of determining the CUEP directly,
we propose a method that can indirectly calculate the mode of instability
without requiring prior knowledge of the CUEP.
At the closest boundary parameter value $p^*$ to $p_0$, the IC lies in the
stable manifold of the CUEP $x^{u}\left(p^{*}\right)$, and thus the system
trajectory converges to $x^{u}\left(p^{*}\right)$.
As $p$ approaches $p^{*}$ from within the set of
recovery values, the system trajectory spends a very large amount
of time around the CUEP $x^{u}\left(p\right)$ before converging to
$x^{s}\left(p\right)$. 
Thus, since the Jacobian depends continuously on state and parameter values,
it seems natural to approximate $A\left(p^{*}\right)$ (the Jacobian
at $x^u\left(p^*\right)$) by averaging the Jacobian along the system trajectory for
$p$ near $p^{*}$.

Unfortunately, averaging the Jacobian along the
system trajectory for all time does not approximate $A\left(p^{*}\right)$ well
since as time approaches infinity, the system converges to
$x^{s}\left(p\right)$. This implies that the Jacobian along the trajectory
converges to the Jacobian at $x^{s}\left(p\right)$, and the average over time
of these Jacobians would also converge to the Jacobian at $x^{s}\left(p\right)$
rather than $A\left(p^*\right)$.
To avoid this, we instead only average the Jacobians along the system trajectory
for a finite length of time $t\left(p\right)$ given by the supremum of the times
at which these Jacobians are unstable.
More precisely, we define
\begin{equation}
\label{eq:4}
t\left(p\right):=\displaystyle\sup_{\substack{t>0\\\frac{\partial f}{\partial x}\left(\phi\left(x_{0}\left(p\right),p,t\right)\right)\text{ is unstable}}}t,
\end{equation}
where $\frac{\partial f}{\partial x}\left(\phi\left(x_{0}\left(p\right),p,t\right)\right)$ is the Jacobian along the system trajectory at time $t$ for parameter
value $p$.

Note that for any $p$ within the set of recovery values,
$t\left(p\right)$ is finite since the system trajectory converges to $x^s\left(p\right)$, and the
Jacobian is stable in a neighborhood of $x^s\left(p\right)$.
The introduction of this $t\left(p\right)$ ensures that the values of the Jacobian
near the CUEP dominate the average of the Jacobians along the system trajectory,
and that this average will converge to $A\left(p^*\right)$ as $p$ approaches $p^*$.
To make this precise, define the piecewise function
$F:R\cup\partial R\rightarrow\mathbb{R}^{n\times n}$
such that for any $p\in R$, $F$ averages the Jacobians along the
system trajectory from the IC up until $t\left(p\right)$, i.e.,
\begin{equation}
\label{eq:5}
F\left(p\right):=\frac{1}{t\left(p\right)}\displaystyle\int_{0}^{t\left(p\right)}{\frac{\partial
    f}{\partial
    x}\left(\phi\left(x_{0}\left(p\right),p,t\right)\right)\,dt}.
\end{equation}
For any $p^{*}\in\partial R$, $t\left(p^{*}\right)=\infty$ since
$x_{0}\left(p^{*}\right)\in
W^{s}\left(x^{u}\left(p^{*}\right)\right)$. Thus, \labelcref{eq:5} is
not defined at $p^{*}$, and so for any $p^{*}\in\partial R$ we define 
\begin{equation}
\label{eq:6}
F\left(p^{*}\right):=A\left(p^{*}\right).
\end{equation}

Theorem \labelcref{the:1} justifies the proposed method for computing
$A\left(p^{*}\right)$ indirectly. It shows that $F\left(p\right)$ will
converge to $F\left(p^{*}\right)$ as $p$ approaches $p^{*}$ from within $R$,
which implies that the average of the Jacobians along the system trajectory
computed in \labelcref{eq:5} converges to $A\left(p^*\right)$.

\begin{theorem}
\label{the:1}
\textit{Assume that $p_0$ satisfies the conditions of Proposition~\ref{pro:2},
  and that Assumptions
  \labelcref{ass:1}-\labelcref{ass:2} hold. Then the piecewise function $F$
  defined in \labelcref{eq:5}-\labelcref{eq:6} is continuous at $p^*$:
\begin{equation}
\label{eq:7}
\displaystyle\lim_{p\rightarrow p^{*}}F\left(p\right)=F\left(p^{*}\right)=A\left(p^{*}\right).
\end{equation}}
\end{theorem}

Theorem \labelcref{the:2} shows that as $p$ approaches $p^{*}$ from within $R$, $F\left(p\right)$ has a unique unstable eigenvalue that converges to the unique unstable eigenvalue of $A\left(p^*\right)$, and its corresponding unstable eigenvector converges to the mode of instability.
\begin{theorem}
\label{the:2}
\textit{Assume the conditions of Theorem \labelcref{the:1}. Then for $p$
  sufficiently close to $p^*$ in $R$, there
  exists a unique unstable eigenvalue $\lambda\left(p\right)$ of
  $F\left(p\right)$, we denote its corresponding unstable eigenvector
  as $v\left(p\right)$, and
\begin{subequations}
\begin{align}
\displaystyle\lim_{p\rightarrow p^{*}}\lambda\left(p\right)&=\lambda\left(p^{*}\right)\label{eq:8a},\\
\displaystyle\lim_{p\rightarrow p^{*}}v\left(p\right)&=v\left(p^{*}\right)\label{eq:8b}.
\end{align}
\end{subequations}}
\end{theorem}

\subsection{Algorithm for Numerically Computing Mode of Instability} \label{sec:algo}

The results of Section~\ref{sec:ave}, and in particular the computation of the
average of the Jacobians in \labelcref{eq:5}, requires exact evaluation of an
integral along the system trajectory.
However, determining the exact system trajectory in continuous time is in
general intractable, and so this section extends the results of
Section~\ref{sec:ave} to the practical setting where the system trajectory
is approximated via numerical integration.
To do so, we will first require a discrete time analog to
Theorems~\ref{the:1}-\ref{the:2}.
Let $h > 0$ represent a constant time step. For any integer $n$, let
$T^n\left(x_0\left(p\right),p,h\right) := \phi\left(x_0\left(p\right),p,nh\right)$ be the state obtained from the continuous
time system \labelcref{eq:1}-\labelcref{eq:2} after $n$ time steps starting from IC
$x_0\left(p\right)$. Analogously to the definition of $t\left(p\right)$ in continuous time,
define $j\left(p\right)$ in discrete time as
\begin{equation}
\label{eq:9}
j\left(p\right):=\displaystyle\sup_{\substack{n\in\mathbb{N},~n > 0\\\frac{\partial f}{\partial x}\left(T^n\left(x_{0}\left(p\right),p,h\right)\right)\text{ is unstable}}} n.
\end{equation}
Finally, analogously to the definition of $F\left(p\right)$ in continuous time,
we define its discrete approximation $\hat{F}\left(p\right)$ by
\begin{equation}
\label{eq:10}
\hat{F}\left(p\right):=\frac{1}{j\left(p\right)}\displaystyle
\sum_{n=0}^{j\left(p\right)}\frac{\partial
  f}{\partial x}\left(T^{n}\left(x_{0}\left(p\right),p,h\right)\right).
\end{equation}

Theorem \labelcref{the:3} shows that as $p$ approaches $p^{*}$
from within $R$, 
$\hat{F}\left(p\right)$ converges
to $A\left(p^{*}\right)$, and it has a unique unstable eigenvalue that converges to the unique unstable eigenvalue of $A\left(p^*\right)$, and its corresponding unstable eigenvector converges to the mode of instability.

\begin{theorem}
\label{the:3}
\textit{Assume the conditions of Theorem \labelcref{the:1}, and fix any
    time step $h > 0$. Then for $p$
  sufficiently close to $p^{*}$ in $R$, there exists a unique unstable
  eigenvalue $\hat{\lambda}\left(p\right)$ of $\hat{F}\left(p\right)$,
  let $\hat{v}\left(p\right)$ denote its corresponding unstable eigenvector,
  and
\begin{subequations}
\begin{align}
\displaystyle\lim_{p\rightarrow p^{*}}\hat{F}\left(p\right)&=A\left(p^{*}\right)\label{eq:11a},\\
\displaystyle\lim_{p\rightarrow p^{*}}\hat{\lambda}\left(p\right)&=\lambda\left(p^{*}\right)\label{eq:11b},\\
\displaystyle\lim_{p\rightarrow p^{*}}\hat{v}\left(p\right)&=v\left(p^{*}\right)\label{eq:11c}.
\end{align}
\end{subequations}}
\end{theorem}

To derive practical guarantees for our approach, fix any autonomous
constant step size numerical integration
scheme with the property that for any IC $x_0$ and any time $t > 0$, the system
state obtained from numerical integration after $n$ time steps
(where $n = \inf_{\substack{n \in \mathbb{N} \\ nh \geq t}} n$)
converges to $\phi\left(x_0,p,t\right)$ as the time step $h$ approaches $0$.
The approximation error resulting from numerical integration can
change the recovery boundary and, thereby, change the value of $p^*$, so
let $p^*\left(h\right)$ denote the unique closest boundary
parameter value on the recovery boundary to $p_0$ for the chosen integration
scheme and the step size $h$.
Define $\Tilde{j}\left(p\right)$ to be the value of $j\left(p\right)$ obtained from
\labelcref{eq:9} when $T^n$ is approximated using numerical integration.
Similarly, define $\tilde{F}\left(p\right)$ to be the value of $\hat{F}\left(p\right)$ obtained
from \labelcref{eq:10} when $T^n$ is approximated using numerical integration. For $h$ sufficiently small, there exist $x^s_h\left(p\right)$ near $x^s\left(p\right)$ which is the hyperbolic SEP, and $x^u_h\left(p\right)$ near $x^u\left(p\right)$ which is the hyperbolic unstable equilibrium point, when $T^n$ is approximated using numerical integration, Theorem \labelcref{the:4} shows that as $h$ approaches $0$,
$p^{*}\left(h\right)$ converges to $p^{*}$. As $p$ approaches $p^*\left(h\right)$ from within the set of recovery values, $\tilde{F}\left(p\right)$ converges to the Jacobian at $x^u_h\left(p^*\left(h\right)\right)$. As $h$ approaches zero, $\tilde{F}\left(p^*\left(h\right)\right)$ converges to $A\left(p^*\right)$, and it has a unique unstable eigenvalue that converges to the unique unstable eigenvalue of $A\left(p^*\right)$, and its corresponding unstable eigenvector converges to the mode of instability.


\begin{theorem}
\label{the:4}
\textit{Assume the conditions of Theorem \labelcref{the:1}.
  Then for $h>0$ sufficiently small,
\begin{subequations}
\begin{align}
  \displaystyle\lim_{h\rightarrow 0}p^{*}\left(h\right)&=p^{*}\label{eq:12a},\\
  \lim_{p \to p^*\left(h\right)} \tilde{F}\left(p\right) &= \frac{\partial f}{\partial x}\left(x^u_h\left(p^*\left(h\right)\right)\right)
  \label{eq:12b},\\
  \lim_{h \to 0}
  \lim_{p \to p^*\left(h\right)} \tilde{F}\left(p\right) &= A\left(p^*\right)
  \label{eq:12c},
\end{align}
\end{subequations}
and there exists a unique unstable eigenvalue
$\Tilde{\lambda}\left(p\right)$ of
$\Tilde{F}\left(p\right)$, with corresponding unstable eigenvector $\Tilde{v}\left(p\right)$, and
\begin{subequations}
\begin{align}
\displaystyle\lim_{h\rightarrow 0}\lim_{p\rightarrow p^*\left(h\right)}\tilde{\lambda}\left(p\right)&=\lambda\left(p^{*}\right)\label{eq:13a},\\
\displaystyle\lim_{h\rightarrow 0}\lim_{p\rightarrow p^*\left(h\right)}\tilde{v}\left(p\right)&=v\left(p^{*}\right)\label{eq:13b}.
\end{align}
\end{subequations}}
\end{theorem}

\section{Numerical Examples}\label{sec:ex}

\subsection{Damped, Driven Nonlinear Pendulum Example}

To illustrate the algorithm of Section~\ref{sec:algo} and
the results of Theorem~\ref{the:4},
we first consider the simple example of
a damped, driven nonlinear pendulum with constant driving force.
For this simple example, it is straightforward to compute the mode of
instability analytically by directly finding the CUEP,
so this example can be used as a benchmark to validate the proposed algorithm.
The dynamics are given by
\begin{align}
\dot{x}_{1}&=x_{2},\label{eq:14},\\
\dot{x}_{2}&=-c_{1}\sin{\left(x_{1}\right)}-c_{2}x_{2}+c_{3}\label{eq:15},
\end{align}
where $c_{1},c_{2},c_{3}>0$ are real parameters and
$x=\left(x_{1},x_{2}\right)\subset\mathbb{R}^{2}$. Physically, $x_{1}$
represents the angle of the pendulum, $x_{2}$ its angular velocity,
$c_{1}$ the square of the natural frequency of the pendulum (under the
small angle approximation),
$c_{2}$ a damping coefficient due to air
drag, and $c_{3}$ the constant driving
torque. Using the simple power system model of a single machine infinite bus \cite{machowski1997power},
\labelcref{eq:14}-\labelcref{eq:15} can also be interpreted as
an electrical generator, with $\left(x_{1},x_{2}\right)$ the angle and
angular velocity of the turbine, $c_{1}$ a constant determining the
electrical torque supplied by the generator, $c_{2}$ a damping
coefficient,
and
$c_{3}$ the constant driving mechanical torque.

For the demonstration below, we set
$c=\left(c_{1},c_{2},c_{3}\right)=\left(2,0.5,1.5\right)$
This system
possesses one SEP at $\left(0.848,0\right)$ and one UEP at
$\left(2.294,0\right)$.
As we vary the parameter values $c_1$, $c_2$, and/or $c_3$, the SEP and UEP
will also vary.

We establish an IC to
\labelcref{eq:14}-\labelcref{eq:15} as the output of the following related
system which models a finite-time disturbance:
\begin{align}
\dot{z}_{1}&=z_{2}\label{eq:16},\\
\dot{z}_{2}&=-c_{2}z_{2}+c_{3}\label{eq:17},
\end{align}
starting from the SEP and integrating for time
$t=0.8$ seconds, which is the
length of time the disturbance is active.
This yields the post-disturbance IC for
\labelcref{eq:14}-\labelcref{eq:15}.
If \labelcref{eq:14}-\labelcref{eq:15} are
interpreted as an electrical generator, then
\labelcref{eq:16}-\labelcref{eq:17} represent a temporary short circuit across
the terminals of the generator, which is modeled by setting the electrical
  torque term in \labelcref{eq:14}-\labelcref{eq:15} to zero temporarily, resulting
  in \labelcref{eq:16}-\labelcref{eq:17}.

For simplicity, consider the one-dimensional parameter space consisting of just the single
parameter $p = c_3$ which we will subsequently treat as a free parameter, with all other parameters constant,
and let $p_0 = 1.5$.
We set the time step $h = 0.02$ seconds, use trapezoidal integration for the numerical integration, and solve each trapezoidal integration
step using the Newton-Raphson method with a tolerance of $10^{-15}$.
Then the numerical integration of \labelcref{eq:16}-\labelcref{eq:17} yields the IC, and the numerical integration of \labelcref{eq:14}-\labelcref{eq:15} provides the full system trajectory, which is used to calculate $\tilde{F}\left(p\right)$ using \labelcref{eq:10}.

\begin{figure}[ht]
\centering
\vspace{-1.4 cm}
\includegraphics[width=1\columnwidth]{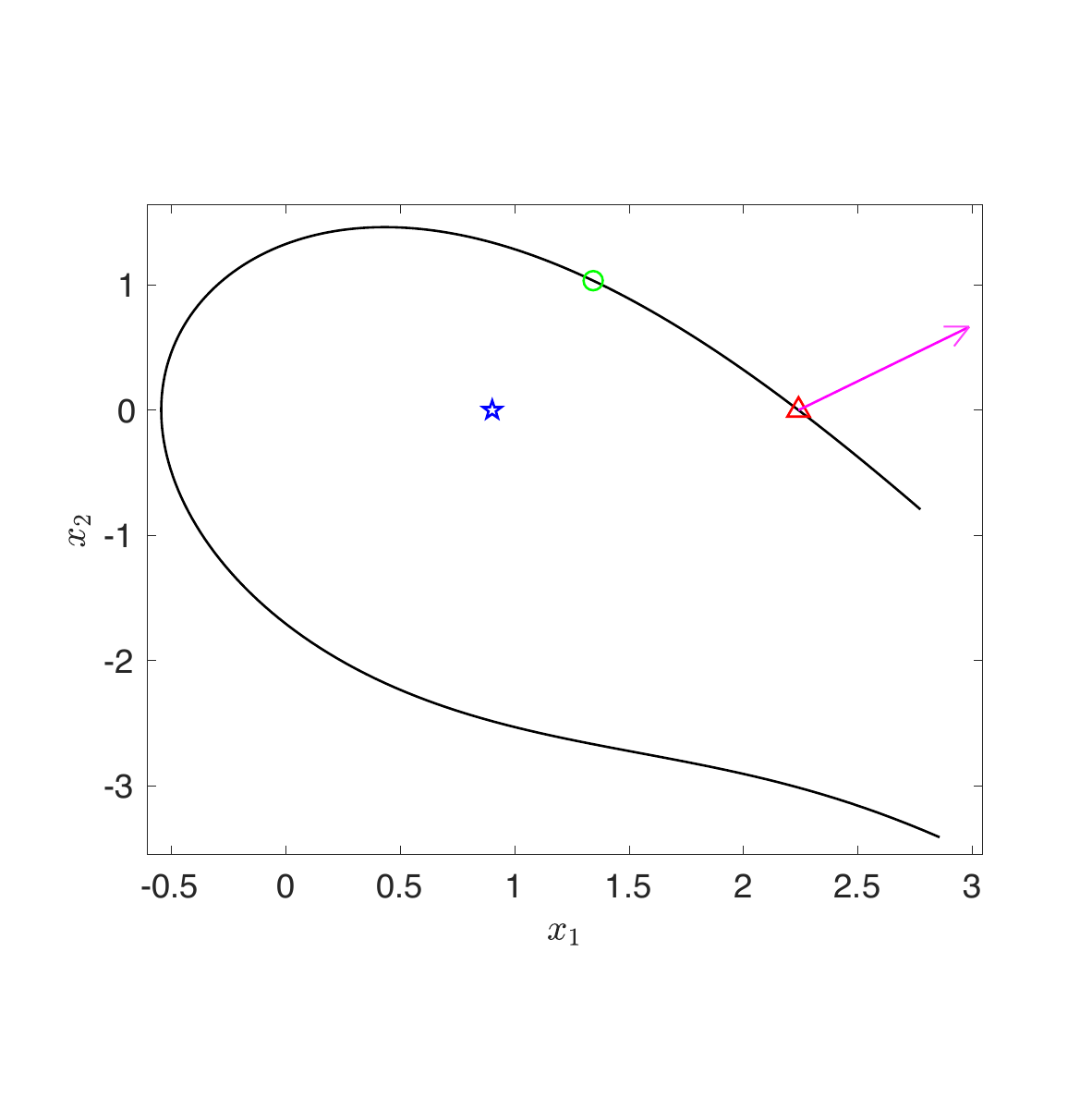}
\vspace{-1.4 cm}
\caption{The region of attraction boundary $\partial W^s\left(x^s_{0.02}\left(\hat{p}\right)\right)$ (solid black line) of the SEP $x^s_{0.02}\left(\hat{p}\right)$ (blue star) of equations \labelcref{eq:16}-\labelcref{eq:17} is shown. It is equal to $W^s\left(x^u_{0.02}\left(\hat{p}\right)\right)$ where $x^u_{0.02}\left(\hat{p}\right)$ (red triangle) is the UEP that can be shown to be the CUEP for the given disturbance. The IC $x_0\left(\hat{p}\right)$ (green circle) is shown. The unstable eigenvector $v\left(\hat{p}\right)$ of $\frac{\partial f}{\partial x}\left(x_{0.02}^u\left(\hat{p}\right)\right)$ (solid magenta line with an arrow), which is the mode of instability for the disturbance, is shown.}

\label{fig:1}
\end{figure}

Since the parameter space is one-dimensional in this case, the closest
boundary parameter value $p^*\left(h\right)$ 
to $p_0$ can be found using a backtracking line
search performed by bisection. 
Due to finite numerical precision, and since $\partial R$ is a set of
measure zero, in general we cannot numerically compute $p^*\left(h\right)$ which
lies exactly on $\partial R$.
Instead, we numerically approximate $p^*\left(h\right)$ to arbitrary accuracy with a value
that lies just inside $R$.
For the remainder of Section~\ref{sec:ex}, we therefore abuse notation and let
$p^*\left(h\right)$ denote this numerical approximation, which lies slightly inside $R$,
rather than the exact theoretical boundary value.
Using the bisection method, we determine that
$p^*\left(0.02\right) = 1.5686$.
In this case, the system with $p = p^*\left(0.02\right)$ has two equilibrium points
(where the angle is restricted to a single interval of length $2\pi$
since the angle is defined modulo $2\pi$), one of which is the SEP
where the system begins prior to the disturbance, while the other is a
UEP that can be shown to be the CUEP for this disturbance
(see Fig.~\ref{fig:1}), consistent with Assumption~\ref{ass:3}.

\begin{figure}
\begin{minipage}[t]{0.475\columnwidth}
  \vspace{-0.5 cm}
  \includegraphics[width=\linewidth]{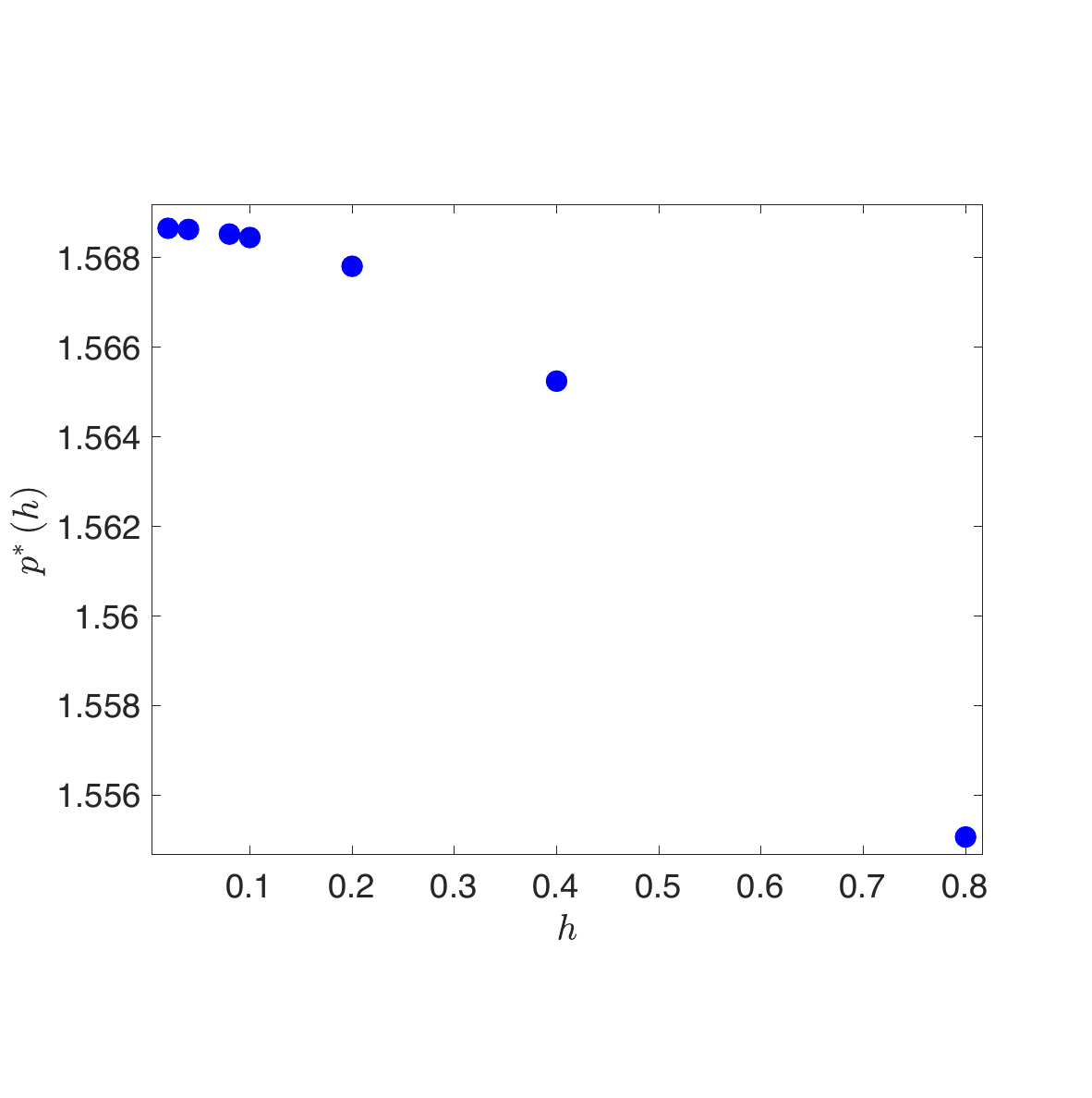}
  \vspace{-1 cm}
  \caption{The closest boundary parameter value $p^*\left(h\right)$ to $p_0$ as a function of $h$. 
  }
  \label{fig:2}
\end{minipage}\hfill 
\begin{minipage}[t]{0.475\columnwidth}
  \vspace{-0.5 cm}
  \includegraphics[width=\linewidth]{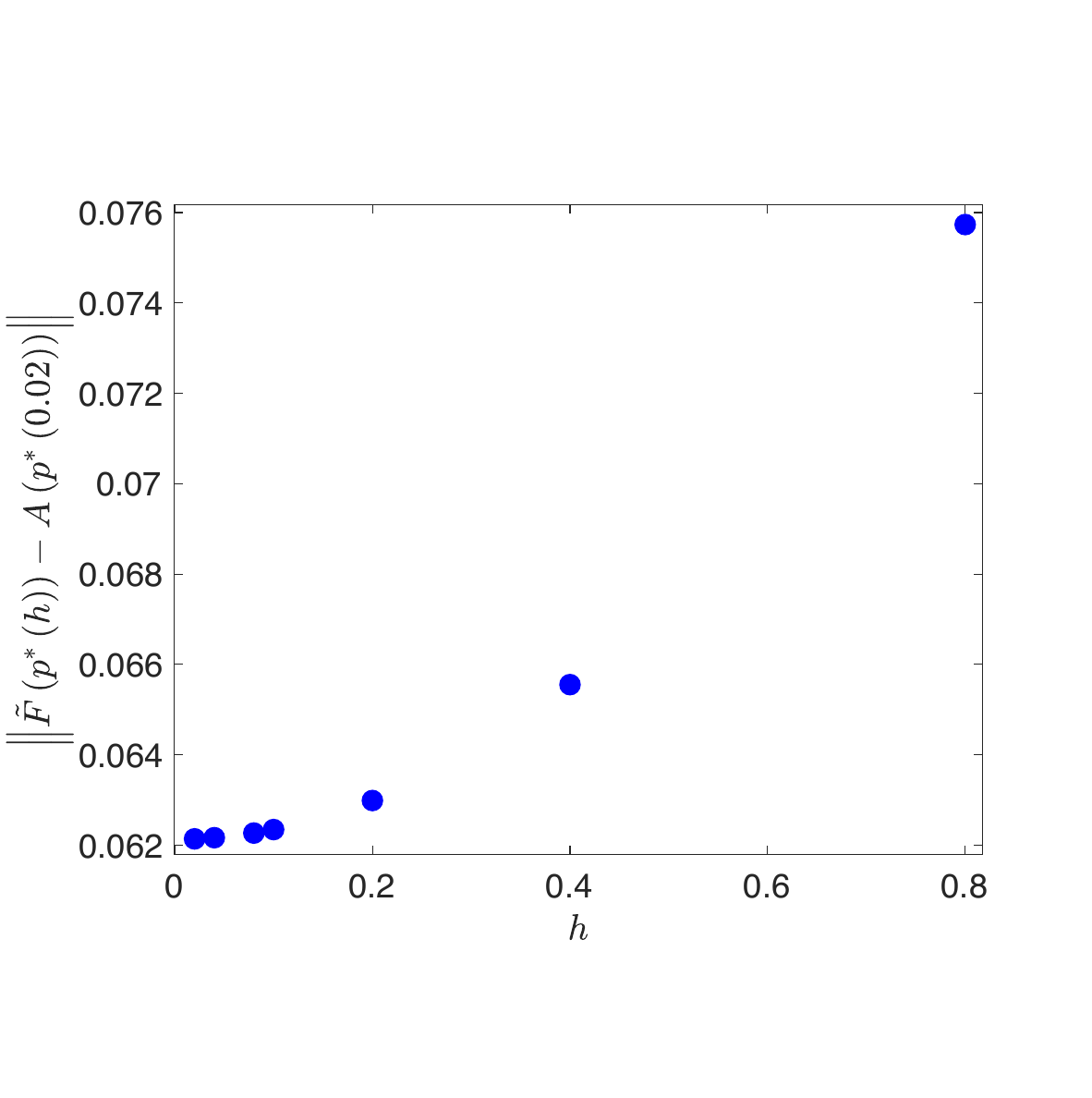}
  \vspace{-1 cm}
  \caption{The norm of the difference between $\tilde{F}\left(p^*\left(h\right)\right)$ and $A\left(p^*\left(0.02\right)\right)$ as a function of $h$. 
  }
  \label{fig:3}
\end{minipage}
\begin{minipage}[t]{0.475\columnwidth}
  \includegraphics[width=\linewidth]{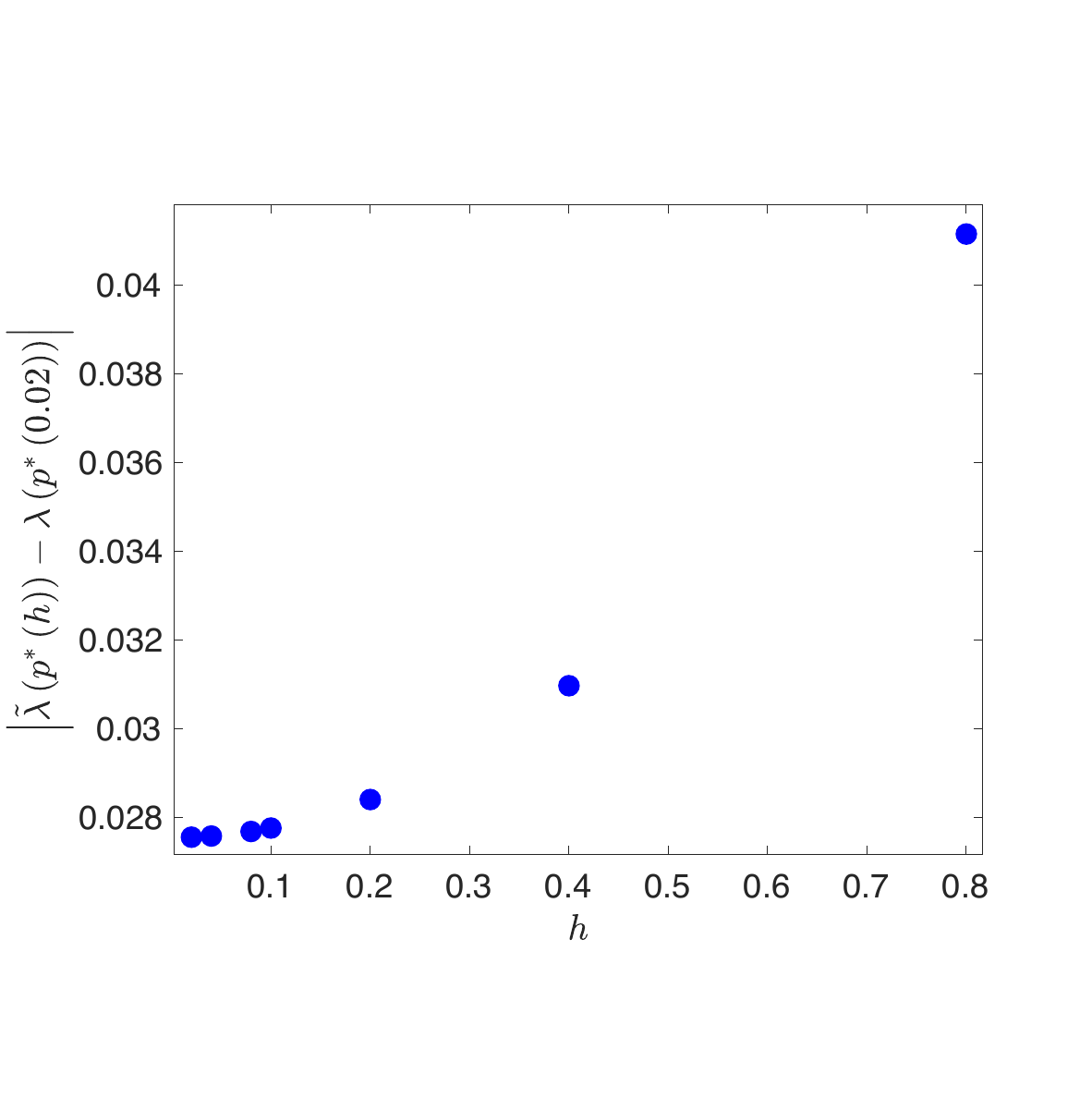}
  \vspace{-1 cm}
  \caption{The absolute value of the difference between the unstable eigenvalues $\tilde{\lambda}\left(p^*\left(h\right)\right)$ of $\tilde{F}\left(p^*\left(h\right)\right)$ and $\lambda\left(p^*\left(0.02\right)\right)$ of $A\left(p^*\left(0.02\right)\right)$ as a function of $h$. 
  }
  \label{fig:4}
\end{minipage}\hfill 
\begin{minipage}[t]{0.475\columnwidth}
  \includegraphics[width=\linewidth]{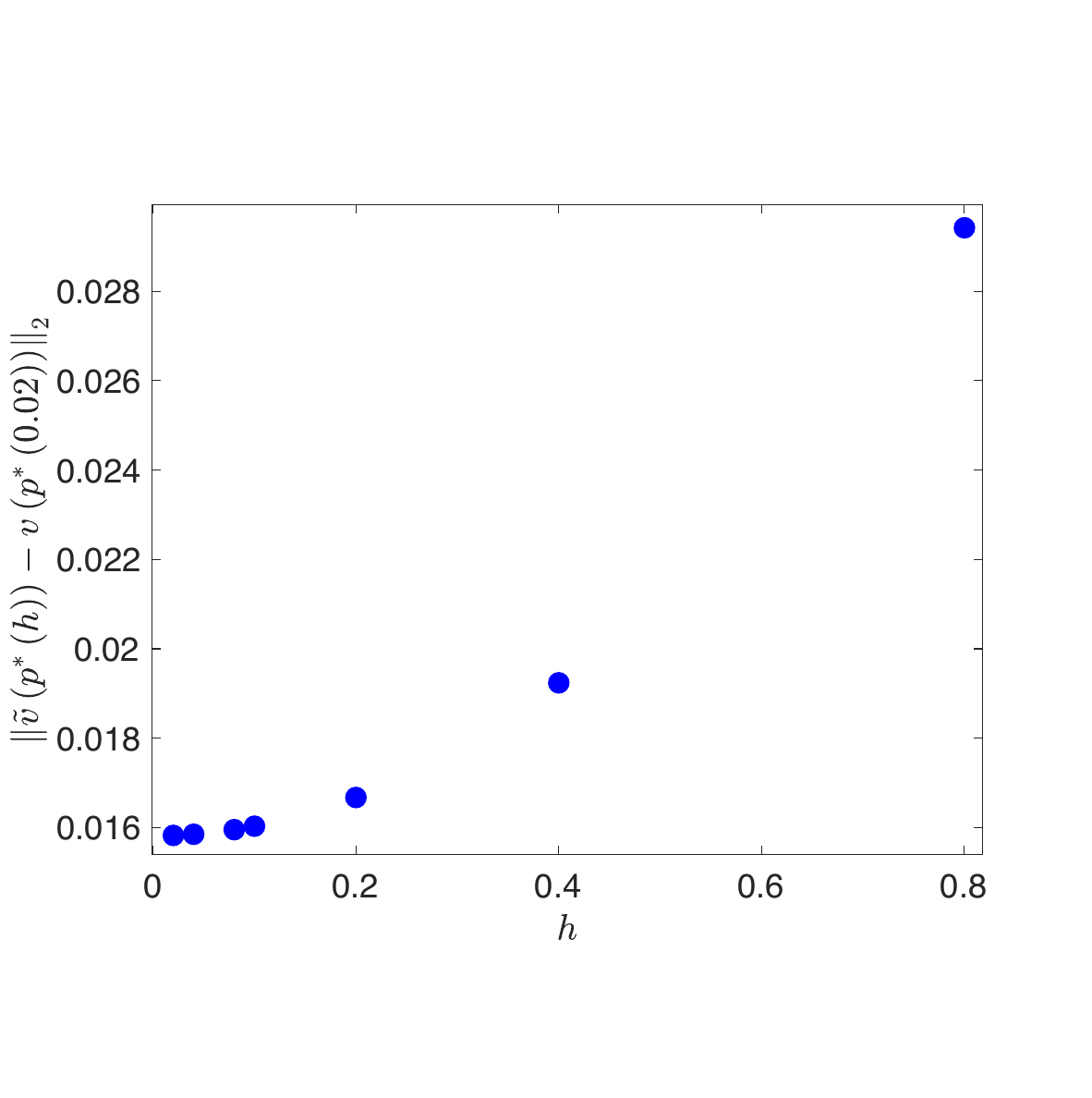}
  \vspace{-1 cm}
  \caption{The norm of the difference between the unstable eigenvectors $\tilde{v}\left(p^*\left(h\right)\right)$ of $\tilde{F}\left(p^*\left(h\right)\right)$ and $v\left(p^*\left(0.02\right)\right)$ of $A\left(p^*\left(0.02\right)\right)$ as a function of $h$. 
  }
  \label{fig:5}
\end{minipage}
\end{figure}

Fig. \labelcref{fig:2} shows the value of $p^*\left(h\right)$ for different values of $h$: $h\in\left\{0.02,0.04,0.08,0.1,0.2,0.4,0.8\right\}$. The graph of $p^*\left(h\right)$ versus $h$ becomes a flat line as $h$ gets sufficiently small, which shows that $p^*\left(h\right)$ converges to the constant $p^*$ as $h$ approaches zero, consistent with \labelcref{eq:12a}. Let $p^*\left(0.02\right)$ be the approximation of the exact theoretical boundary parameter value. For the same values of $h$, $\left\|\tilde{F}\left(p^*\left(h\right)\right)-\frac{\partial f}{\partial x}\left(x_h^u\left(p^*\left(h\right)\right)\right)\right\|<6.7\%$, which is consistent with \labelcref{eq:12b}. Fig. \labelcref{fig:3} shows the value of $\left\|\tilde{F}\left(p^*\left(h\right)\right)-A\left(p^*\left(0.02\right)\right)\right\|$ for the same values of $h$, which shows that $\tilde{F}\left(p^*\left(h\right)\right)$ converges to the true Jacobian as $h$ approaches zero, consistent with \labelcref{eq:12c}. Fig. \labelcref{fig:4} and Fig. \labelcref{fig:5} show $\left|\tilde{\lambda}\left(p^*\left(h\right)\right)-\lambda\left(p^*\left(0.02\right)\right)\right|$ and $\left\|\tilde{v}\left(p^*\left(h\right)\right)-v\left(p^*\left(0.02\right)\right)\right\|_2$ for the same values of $h$, which shows that the unstable eigenvalue and its associated eigenvector of $\tilde{F}\left(p^*\left(h\right)\right)$ converge to the unstable eigenvalue of the true Jacobian and the mode of instability, respectively, as $h$ approaches zero, consistent with \labelcref{eq:13a} and \labelcref{eq:13b}. As the CUEP $x_{0.02}^u\left(p^*\left(0.02\right)\right)$ is known and available analytically, the Jacobian at the CUEP,
$A\left(p^*\left(0.02\right)\right)$, can therefore be computed directly and exactly.
Consistent with 
Assumption~\ref{ass:2}, it has a single unstable eigenvalue, and
its corresponding unstable eigenvector is
$v\left(p^*\left(0.02\right)\right) =\begin{bmatrix} 0.7464\\ 0.6655\end{bmatrix}$.
Let $\hat{p} = p^*\left(0.02\right)$ in this case, and note that $\hat{p} \in R$ and is very
close to the exact theoretical boundary value.
Numerically computing $\tilde{F}\left(\hat{p}\right)$ from the system
trajectory, consistent with Theorem~\ref{the:4}, it is shown to have a single
unstable eigenvalue with associated unstable eigenvector
$\tilde{v}(\hat{p}) =\begin{bmatrix}0.7569\\0.6536\end{bmatrix}$. 
Thus, $\left\|\tilde{v}\left(\hat{p}\right)-v\left(p^*\left(0.02\right)\right)\right\|_2 < 0.016$,
so $\tilde{v}\left(\hat{p}\right)$ and $v\left(p^*\left(0.02\right)\right)$ have less than 1.6\% difference since
both are normalized to unit length.
Fig. \labelcref{fig:1} also shows the region of attraction boundary and the mode of instability at $\hat{p}$. Therefore, this illustrates that the nonlinear mode of instability of this low
dimensional system can be accurately computed by the proposed algorithm.

\subsection{IEEE 9-Bus Power System}

\begin{figure}[ht]
\centering
\includegraphics[width=\columnwidth]{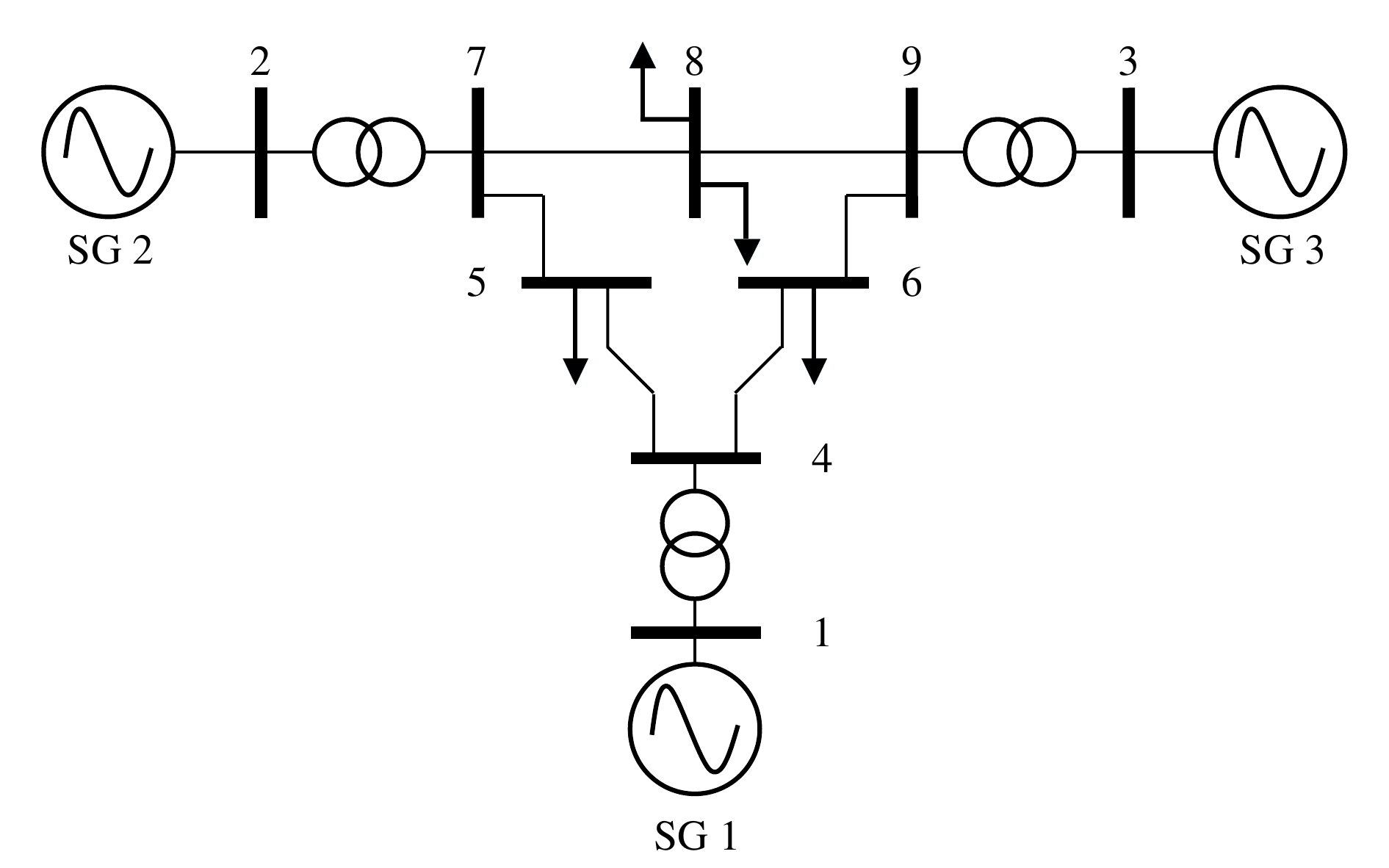}
\vspace{-0.5 cm}
\caption{IEEE 9-Bus Power System (SG = synchronous generator).}
\label{fig:6}
\end{figure}

Next, the algorithm described in Section~\ref{sec:algo}
is applied to the more complex example of the IEEE 9-bus power system \cite{anderson2003the},
where the mode of instability is not so straightforward to obtain analytically.
As shown in Fig.~\ref{fig:6},
this is a network consisting of nine nodes that contains
three synchronous generators, three loads, six transmission lines, and three transformers.
Each generator is modeled using a 4th order synchronous machine model,
along with an Active Voltage Regulator (AVR) and Power System Stabilizer (PSS)
for control, which are modeled according to the IEEE standard ST1C and PSS1A
models \cite{ieee2016ieee}, respectively.
The loads are modeled as constant
power loads.
The dimension of the dynamic states for this system is 60.
The disturbance considered is a temporary short circuit, known as a fault,
at the terminals of generator three.
The system is initially at a SEP, then the fault occurs at 0.02 seconds.
It lasts for 0.2 seconds and is cleared at
0.22 seconds.
For this nonlinear power system model with higher dimensions, the mode of
instability is determined without direct identification of the CUEP
using the algorithm from Section~\ref{sec:algo}.

Many model parameters of the system are of interest for recovery considerations. The moment of inertia scaling factor multiplies the moment of inertia of all three synchronous generators simultaneously. The gains of the AVR controllers can help to capture the impact of controller turning on system stability and the gains of the PSS controllers are crucial for enhancing system stability by improving the damping of low-frequency oscillations in power systems. The active and reactive power loads are modeled using the standard exponential form of a voltage-dependent load model. The voltage exponents are set uniformly for all active power loads and similarly for all reactive power loads. As load is often uncertain, given the complexity of modeling load dynamics, these parameters can help to capture the impact of uncertain load behavior on system recoverability.

To demonstrate the effectiveness of the proposed algorithm, we first
consider the one-dimensional parameter space with the parameter of
interest being the moment of inertia scaling factor. We use
trapezoidal integration with the Newton-Raphson method for the numerical
integration 
with a fixed time step of $h = \f{1}{60}$ seconds and a tolerance of $10^{-10}$.
This choice of $h$ represents a single cycle of the $60$~Hz AC voltage
  and, thus, is faster than all other dynamics explicitly included in the
  system model,
  so it represents a choice of $h$ that is close to zero for this application.
Using the backtracking line search performed by bisection, or using the method from \cite{fisher2025computing},
which finds the closest point on the recovery boundary in one-dimensional parameter space,
we compute the 
boundary value of the moment of inertia scaling factor to be
$p^*\left(\frac{1}{60}\right)=0.8724$, which is very close to the exact theoretical
boundary value.
For this example, we do not know the location of the
CUEP, so we are unable to compute $v\left(p^*\left(\frac{1}{60}\right)\right)$ directly from
$A\left(p^*\left(\frac{1}{60}\right)\right)$.
Instead, we set $\hat{p} = p^*\left(\frac{1}{60}\right)$, and numerically compute
$\tilde{F}\left(\hat{p}\right)$.

Consistent with Theorem~\ref{the:4},
$\tilde{F}\left(\hat{p}\right)$ has exactly one 
unstable eigenvalue $\tilde{\lambda}\left(\hat{p}\right)$.
As the choice of $h$ is close to zero for this application,
by Theorem~\ref{the:4} its corresponding eigenvector $\tilde{v}\left(\hat{p}\right)$
is a close approximation to the true mode of instability.
Table~\ref{tab:1} shows the components of $\tilde{v}\left(\hat{p}\right)$ other than
those which are close to zero.
The components of $\omega_1$, $\omega_2$, and $\omega_3$ of
$\tilde{v}\left(\hat{p}\right)$ correspond to
the frequency components of generator one, generator two, and generator three, respectively. The components of $\theta_1$, $\theta_2$, and $\theta_3$ of $\tilde{v}\left(\hat{p}\right)$ correspond to the angle of generator one, generator two, and generator three, respectively. The components of $x_1$, $x_2$, and $x_3$ correspond to the internal state of the AVR of generator one, generator two, and generator three, respectively.
\begin{table}[ht]
\begin{center}
\begin{tabular}{|c|c|c|}
\hline
$\omega_{1}$&$\omega_{2}$&$\omega_{3}$\\\hline
$-0.0112$&$0.7594$&$0.4227$\\\hline
$\theta_{1}$&$\theta_{2}$&$\theta_{3}$\\\hline
$-0.0026$&$0.1754$&$0.0977$\\\hline
$x_1$&$x_2$&$x_3$\\\hline
$0.0863$&$0.3473$&$0.2673$\\\hline
\end{tabular}
\end{center}
\caption{Components of $\tilde{v}\left(\hat{p}\right)$ representing the frequencies $\omega$, angles $\theta$, and the internal state of the AVR $x$ for the three synchronous generators in the case of one-dimensional parameter space.}
\label{tab:1}
\end{table}

\begin{figure}[ht]
\centering
\vspace{-1.5 cm}
\includegraphics[width=\columnwidth]{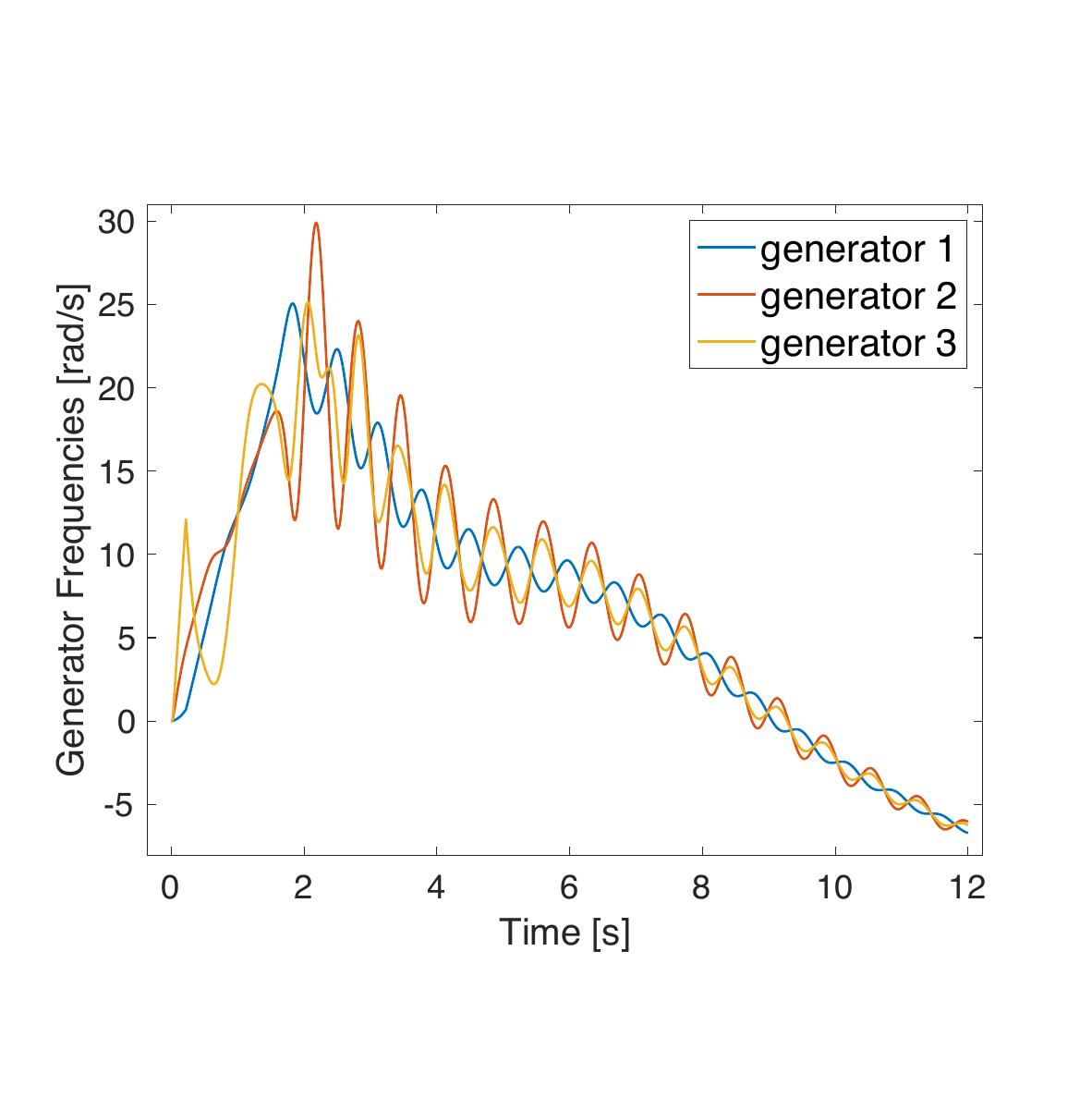}
\vspace{-1.5 cm}
\caption{The frequency dynamics of all three generators before reaching synchronism in the one-dimensional parameter space, where the moment of inertia scaling factor is the parameter of interest, at $p = p^*\left(\frac{1}{60}\right)$.}
\label{fig:7}
\end{figure}

\begin{figure}[ht]
\centering
\vspace{-1.25 cm}
\includegraphics[width=\columnwidth]{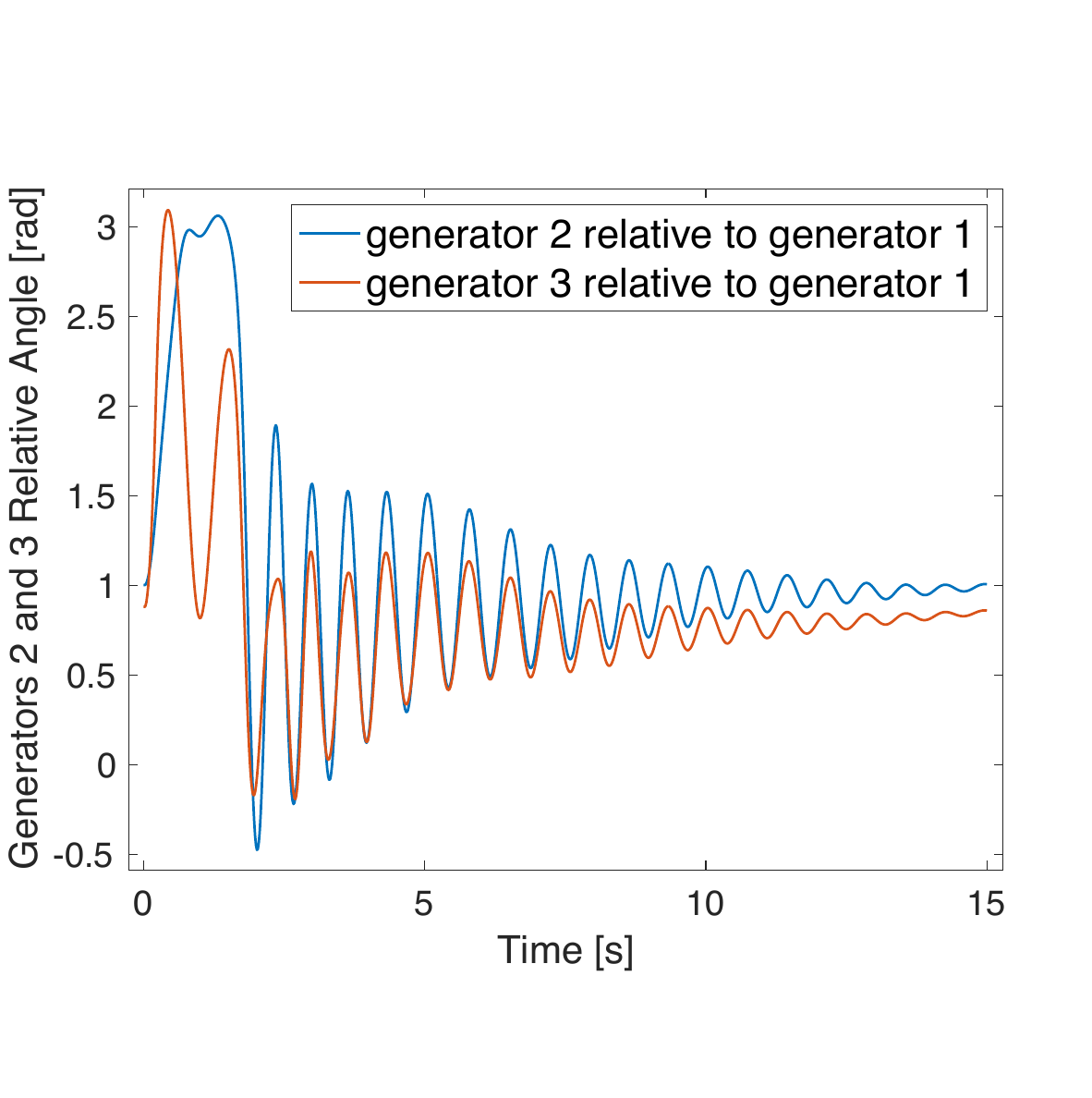}
\vspace{-1.5 cm}
\caption{The angle of generator two and generator three relative to generator one in the one-dimensional parameter space, where the moment of inertia scaling factor is the parameter of interest, at $p=p^*\left(\frac{1}{60}\right)$.}
\label{fig:8}
\end{figure}

\begin{figure}[ht]
\centering
\vspace{-1.4 cm}
\includegraphics[width=\columnwidth]{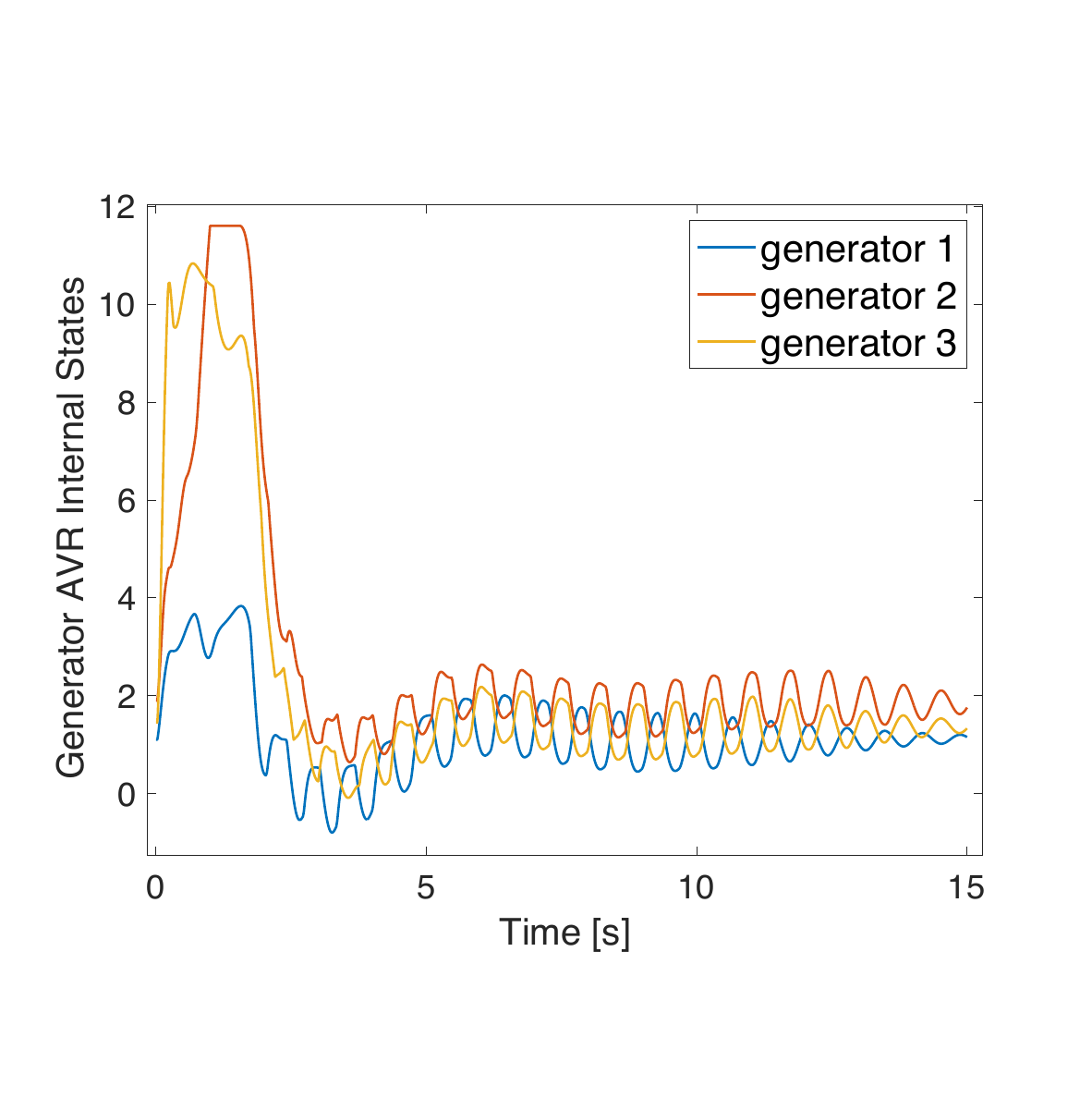}
\vspace{-1.5 cm}
\caption{The internal state dynamics of the AVR of all three synchronous generators in the one-dimensional parameter space, where the moment of inertia scaling factor is the parameter of interest, at $p=p^*\left(\frac{1}{60}\right)$.}
\label{fig:9}
\end{figure}

The mode of instability, as shown in Table~\ref{tab:1},  reveals the subset of dynamics responsible for failure to recover from the fault. In particular, among the three generators, the frequency component, angle component, and the AVR internal state component of generator two are the largest in magnitude. This leads to the non-intuitive information that generator two is most responsible for the failure to recover from the disturbance, even though the disturbance originates at a distant location from generator two (near generator three). To further validate this observation, we plot the frequency dynamics of all three generators before they achieve synchronism, as shown in Fig.~\ref{fig:7}. The figure shows that the frequency of generator two has larger amplitude oscillations after the disturbance has occurred than the frequencies of the other two generators. Fig.~\ref{fig:8} shows the angle of generator three relative to generator one and the angle of generator two relative to generator one as a function of time. We observe that the relative angle of generator two has larger amplitude oscillations after the disturbance has occurred than the relative angle of generator three, indicative of generator two's role as the generator most important to the mechanism of instability for the disturbance. Fig. \labelcref{fig:9} shows the dynamics for the internal state of the AVR of all three generators. It shows that the internal state of the AVR of generator two has a larger initial transient than the internal state of the AVR of generator one and generator three, which indicates that generator two is most important for determining whether the system will fail to recover from the disturbance.
In addition, Table~\ref{tab:1} shows that the frequency component, angle component, and the AVR internal state component of generator one are very small, indicating that generator one plays a more minimal role in the onset of failure to recover than generators two and three. Furthermore, the components of the internal states of the PSS are close to zero, indicating the nontrivial fact that, unlike for many other faults, for this particular disturbance the PSS controllers do not have a significant contribution to the mechanism of instability.

For the multi-dimensional parameter space, we consider the parameter
set consisting of the moment of inertia of each of the three generators,
the real and reactive load voltage exponents for each of the three loads, the
AVR and PSS controller gains for each of the three generators (15 parameters).
This set captures both the
uncertainty in load dynamic characteristics and the influence of
controller response on system recoverability. Let $H_{1}$, $H_{2}$,
and $H_{3}$ be the moment of inertia of generator one, generator two,
and generator three, respectively. Let $e_{1}$, $e_{2}$, and $e_{3}$ be the real load voltage exponents for each of the three loads. Let $e_4$, $e_5$, and $e_6$  be the reactive load voltage exponents for each of the three loads. Let $K_{1}$, $K_{2}$, and $K_{3}$ be the AVR controller gains of generator
one, generator two, and generator three, respectively. Let $L_1$, $L_2$, and $L_3$ be the PSS controller gains of generator one, generator two, and generator three, respectively. Let $H=\begin{bmatrix}H_1&H_2&H_3\end{bmatrix}^{\intercal}$, $e=\begin{bmatrix}e_1&e_2&e_3&e_4&e_5&e_6\end{bmatrix}^{\intercal}$, $K=\begin{bmatrix}K_1&K_2&K_3\end{bmatrix}^{\intercal}$, and $L=\begin{bmatrix}L_1&L_2&L_3\end{bmatrix}^{\intercal}$. Let
$p=\begin{bmatrix}H&e&K&L\end{bmatrix}^{\intercal}$
  be the
vector of chosen parameters. Let $p_{0}$ be the initial parameter values, as
given in Table \labelcref{tab:2}. We apply the optimization approach
\cite{fisher2025computing}
to find the closest point on the recovery
boundary to the nominal value of this 15-dimensional parameter space,
with the solution tolerance set to $\epsilon=10^{-5}$.

\begin{table}[ht]
\begin{center}
\begin{tabular}{|c|c|c|c|c|}
\hline
$H_{1}$&$H_{2}$&$H_{3}$&$e_{1}$&$e_{2}$\\\hline
$\frac{23.64}{60\pi}$&$\frac{6.4}{60\pi}$&$\frac{3.01}{60\pi}$&$2$&$2$\\\hline
$e_{3}$&$e_{4}$&$e_{5}$&$e_{6}$&$K_{1}$\\\hline
$2$&$2$&$2$&$2$&$20$\\\hline
$K_2$&$K_3$&$L_1$&$L_2$&$L_3$\\\hline
$20$&$20$&$\frac{0.5}{120\pi}$&$\frac{0.5}{120\pi}$&$\frac{0.5}{120\pi}$\\\hline
\end{tabular}
\end{center}
\caption{Test case parameter values.}
\label{tab:2}
\end{table}

The parameter value on the recovery boundary $p^*\left(\frac{1}{60}\right)=\begin{bmatrix}H^*&e^*&K^*&L^*\end{bmatrix}^{\intercal}$
for this 15-dimensional parameter space is shown in Table
\labelcref{tab:3},
which closely approximates the exact theoretical boundary value, and is arbitrarily close to the initial parameter value. Since the location of the CUEP is unknown, we are unable to determine $A\left(p^*\left(\frac{1}{60}\right)\right)$ and, consequently, $v\left(p^*\left(\frac{1}{60}\right)\right)$ directly. Instead, we set $\hat{p}=p^*\left(\frac{1}{60}\right)$ once again and numerically compute $\tilde{F}\left(\hat{p}\right)$.
\begin{table}[ht]
\begin{center}
\vspace{0.16 cm}
\begin{tabular}{|c|c|c|c|c|}
\hline
$H_1^*$&$H_2^*$&$H_3^*$&$e_1^*$&$e_2^*$\\\hline
$0.1254$&$0.0340$&$0.0160$&$2$&$2$\\\hline
$e_3^*$&$e_4^*$&$e_5^*$&$e_6^*$&$K_1^*$\\\hline
$2$&$2$&$2$&$2$&$20$\\\hline
$K_2^*$&$K_3^*$&$L_1^*$&$L_2^*$&$L_3^*$\\\hline
$20$&$20$&$0.0013$&$0.0013$&$0.0013$\\\hline
\end{tabular}
\end{center}
\caption{Closest boundary parameter value for moment of inertia, real and reactive load voltage exponents, AVR and PSS controller gains.}
\label{tab:3}
\end{table}

Consistent with Theorem \labelcref{the:4},
$\tilde{F}\left(\hat{p}\right)$ possesses a unique unstable eigenvalue
and, by Theorem~\ref{the:4}, its associated unstable eigenvector is a
close approximation to the mode of instability. Table
\labelcref{tab:4} presents the components of
$\tilde{v}\left(p^*\left(\frac{1}{60}\right)\right)$ apart from those that are close to
zero. The components of $\omega_1$, $\omega_2$, $\omega_3$,
$\theta_1$, $\theta_2$, $\theta_3$, $x_1$, $x_2$, $x_3$ of
$\tilde{v}\left(\hat{p}\right)$ carry the same interpretation as those
in the one-dimensional parameter space.

\begin{table}[ht]
\begin{center}
\begin{tabular}{|c|c|c|}
\hline
$\omega_{1}$&$\omega_{2}$&$\omega_{3}$\\\hline
$0.1106$&$0.4388$&$0.6850$\\\hline
$\theta_{1}$&$\theta_{2}$&$\theta_{3}$\\\hline
$0.0300$&$0.1192$&$0.1861$\\\hline
$x_1$&$x_2$&$x_3$\\\hline
$0.1150$&$0.2648$&$0.4267$\\\hline
\end{tabular}
\end{center}
\caption{Components of $\tilde{v}\left(\hat{p}\right)$ representing the frequencies $\omega$, angles $\theta$, and the internal state of the AVR $x$ for the three synchronous generators in the case of 15-dimensional parameter space.}
\label{tab:4}
\end{table}

\begin{figure}[ht]
\centering
\vspace{-1.5 cm}
\includegraphics[width=\columnwidth]{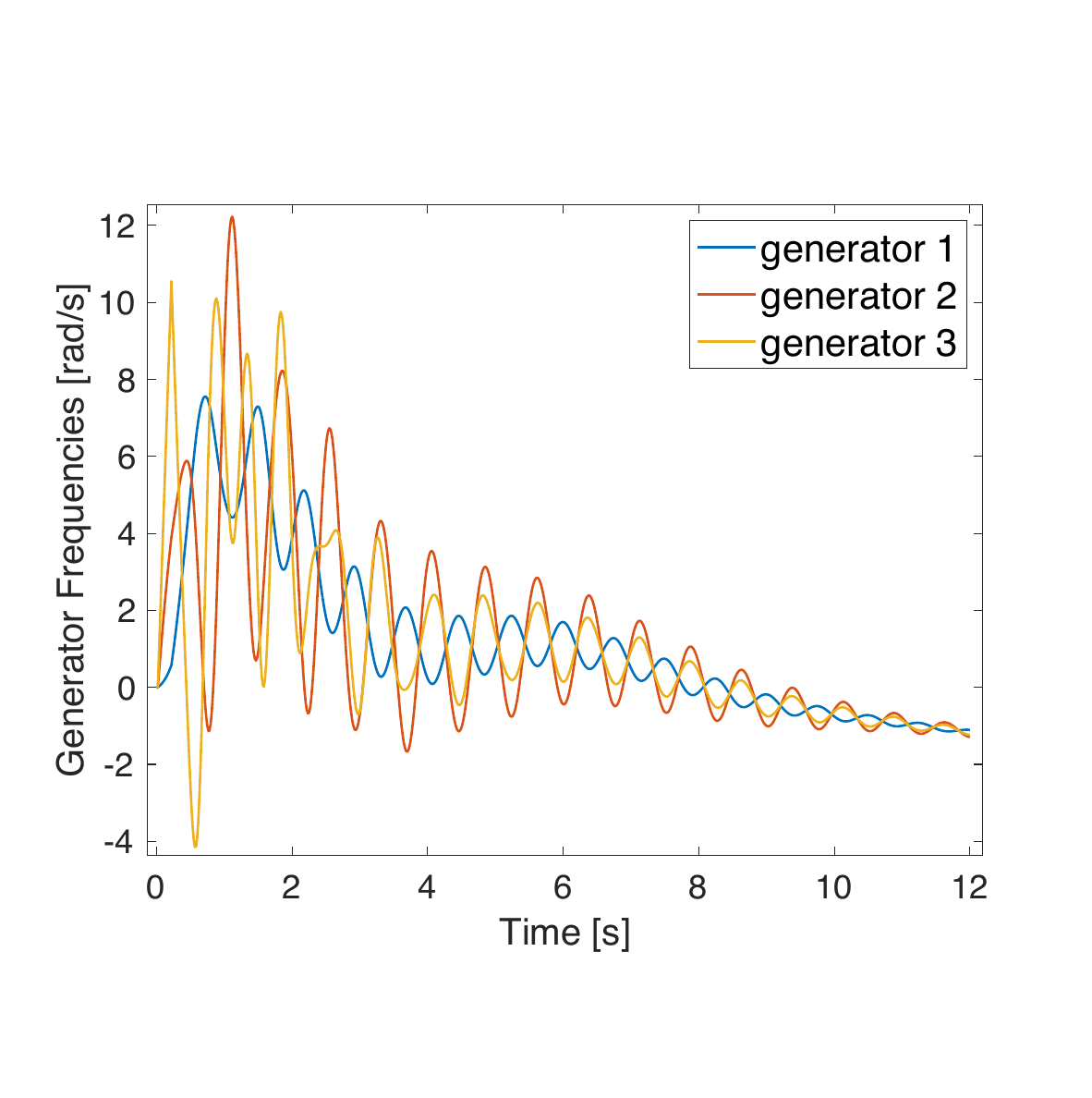}
\vspace{-1.5 cm}
\caption{The frequency dynamics of all three generators before reaching synchronism in the 15-dimensional parameter space, where the parameters of interest are the moment of inertia, real and reactive load voltage exponents, AVR and PSS controller gains, at $p=p^*\left(\frac{1}{60}\right)$.}
\label{fig:10}
\end{figure}

\begin{figure}[ht]
\centering
\vspace{-1.5 cm}
\includegraphics[width=\columnwidth]{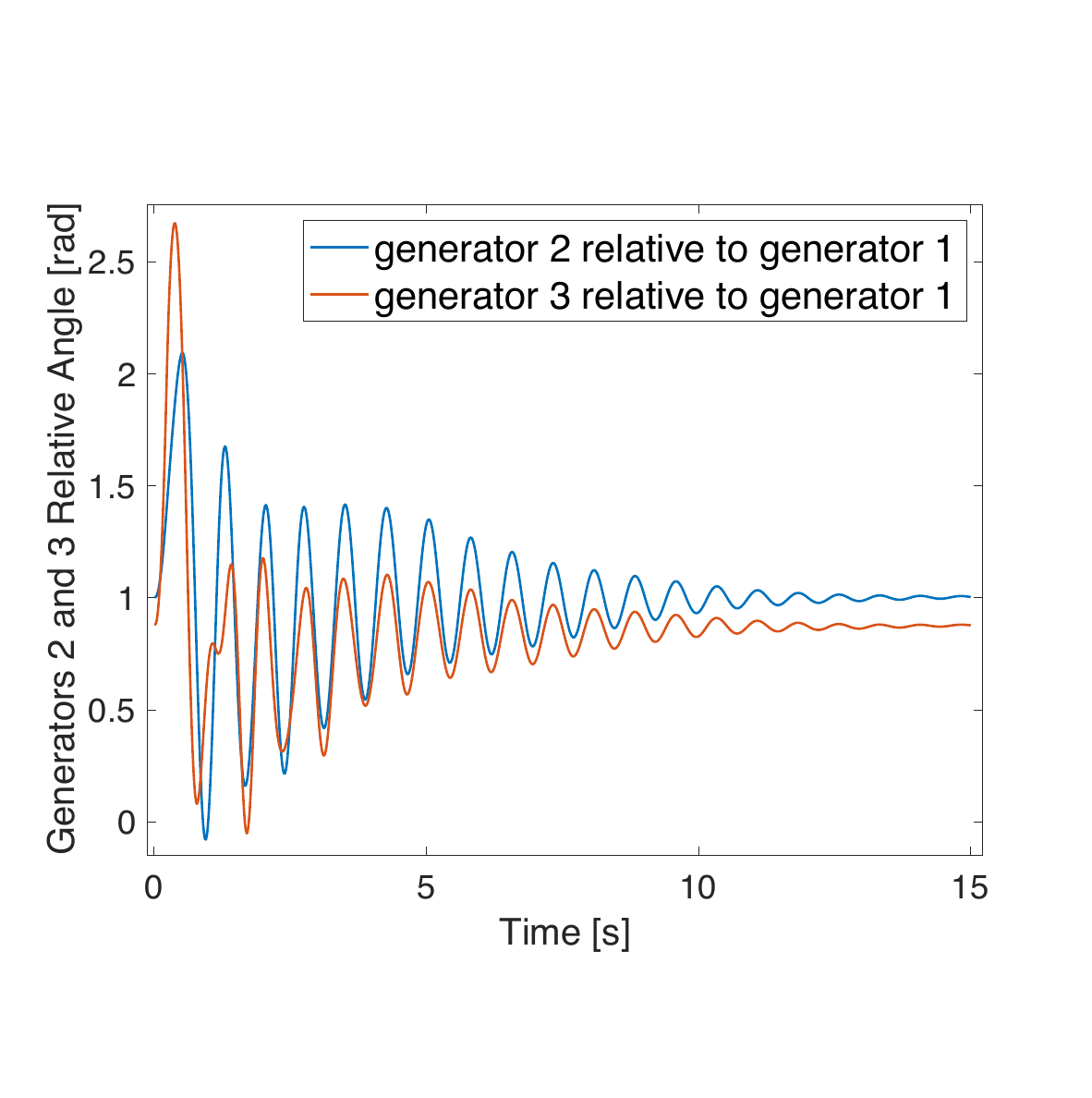}
\vspace{-1.5 cm}
\caption{The angle of generator two and generator three relative to generator one in the 15-dimensional parameter space, where the parameters of interest are the moment of inertia, real and reactive load voltage exponents, AVR and PSS controller gains, at $p=p^*\left(\frac{1}{60}\right)$.}
\label{fig:11}

\end{figure}

\begin{figure}[ht]
\centering
\vspace{-1.4 cm}
\includegraphics[width=\columnwidth]{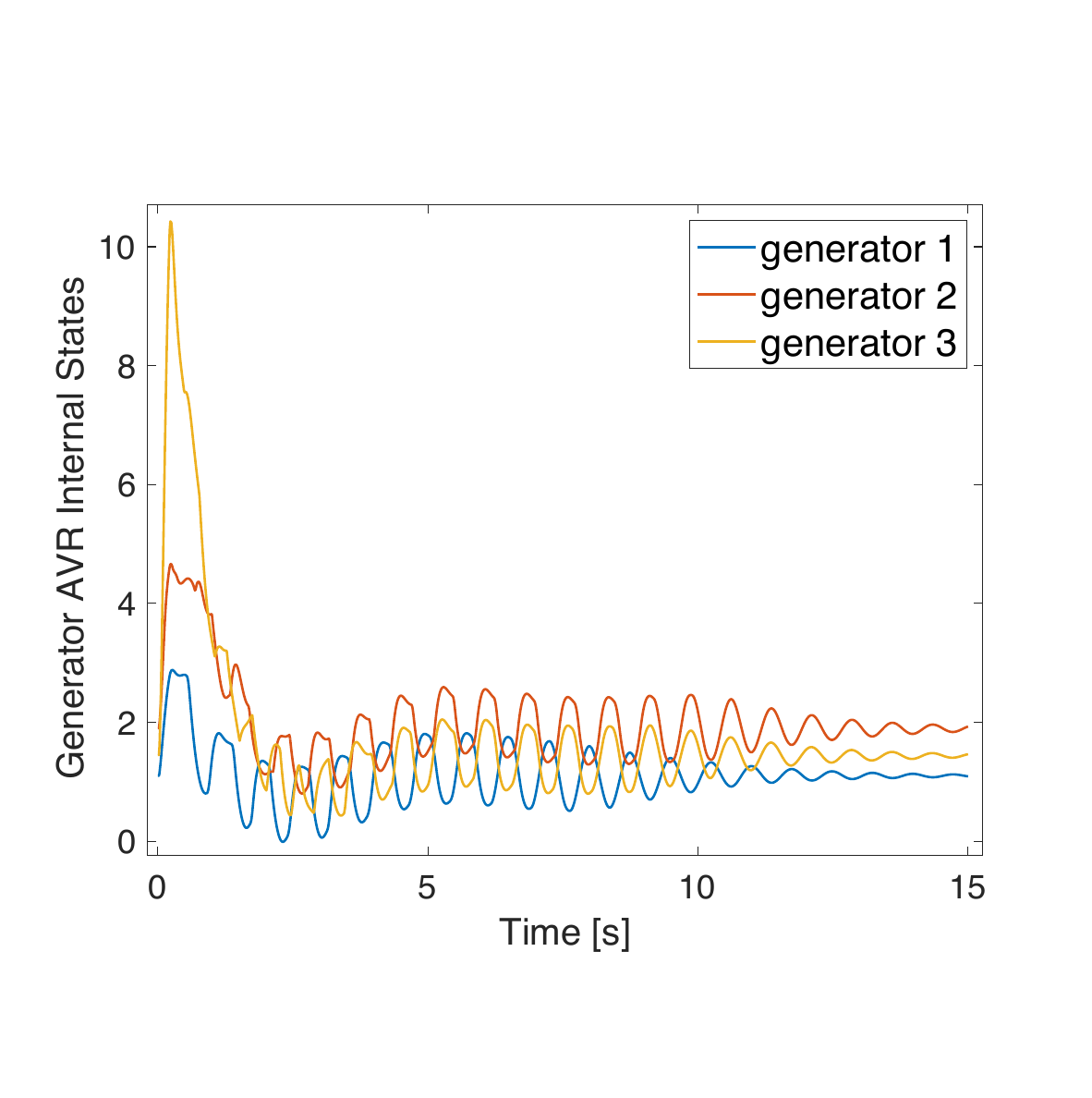}
\vspace{-1.5 cm}
\caption{The internal state dynamics of the AVR of all three synchronous generators in the 15-dimensional parameter space, where the parameters of interest are the moment of inertia, real and reactive load voltage exponents, AVR and PSS controller gains, at $p=p^*\left(\frac{1}{60}\right)$.}
\label{fig:12}
\end{figure}

Unlike the one-dimensional case with the moment of inertia scaling factor, for the given higher dimensional parameter space, generator three has the largest frequency, angle, and the AVR internal state components in magnitude, followed by generator two and then generator one. This provides a clear indication that generator three plays the most significant role in the failure to recover from the disturbance, which is intuitive since the disturbance originates near generator three, followed by generator two and then generator one. To further validate this observation, the frequency dynamics of all three generators before they reach synchronism, and the angle of generator three and generator two relative to generator one as a function of time, are shown in Fig.~\ref{fig:10} and Fig.~\ref{fig:11}, respectively. We observe that after the fault is cleared, the frequency of generator three exhibits a long transient, with large oscillations starting earlier and ending later than for the other generators, with generator two having the second largest initial frequency transient, and finally generator one having the smallest. Furthermore, the relative angle of generator three has greater amplitude oscillations than those of generator two. Fig. \labelcref{fig:12} illustrates the AVR internal state dynamics for all three generators. We can see that the internal state of the AVR of generator three has a larger initial transient than the internal state of the AVR of generator two, and generator one has the smallest initial transient in its AVR internal state.
All of these observations highlight that the frequency and angle dynamics of generator three, along with the operation of its AVR, are the primary factors influencing whether the system can recover from the disturbance, followed by those of generator two, and then of generator one. Notably and non-intuitively, although both the frequency and voltage dynamics play a major role in the mechanism of instability in this case, the dynamics of the PSS, which are designed to limit the coupling between frequency and voltage dynamics, do not play a significant role in the failure to recover.

Overall, the proposed algorithm was successfully applied to the IEEE 9-bus benchmark power system to compute the nonlinear mode of instability for both one-dimensional and higher dimensional parameter spaces, and revealed unexpected dynamic behavior responsible for the mechanism of instability that would not have been straightforward to identify otherwise.

\section{Proofs}\label{sec:proof}
\begin{proof}[Proof of Theorem \labelcref{the:1}]
  This proof proceeds by splitting the system trajectory for $p \in R$
  with $p$ near $p^*$ into three segments as illustrated in Fig.~\ref{fig:13}:
  (i) a segment that starts from the IC $x_0\left(p\right)$ and enters a
  ball $N$ around the CUEP, (ii) another segment that travels within $N$, and
  (iii) a final segment that leaves $N$ and enters the local stable manifold of
  the SEP.  The key idea is that as $p$ approaches $p^*$, the system trajectory
  spends only a finite amount of time on segments (i) and (iii), but the amount
  of time it spends in segment (ii) diverges towards infinity.
  Thus, the value of the Jacobian along segment (ii), which is very close to
  the Jacobian at the CUEP, will dominate in the calculation of the average
  $F\left(p\right)$ as $p$ approaches $p^*$.
  
\begin{figure}[ht]
\centering   \includegraphics[width=\columnwidth]{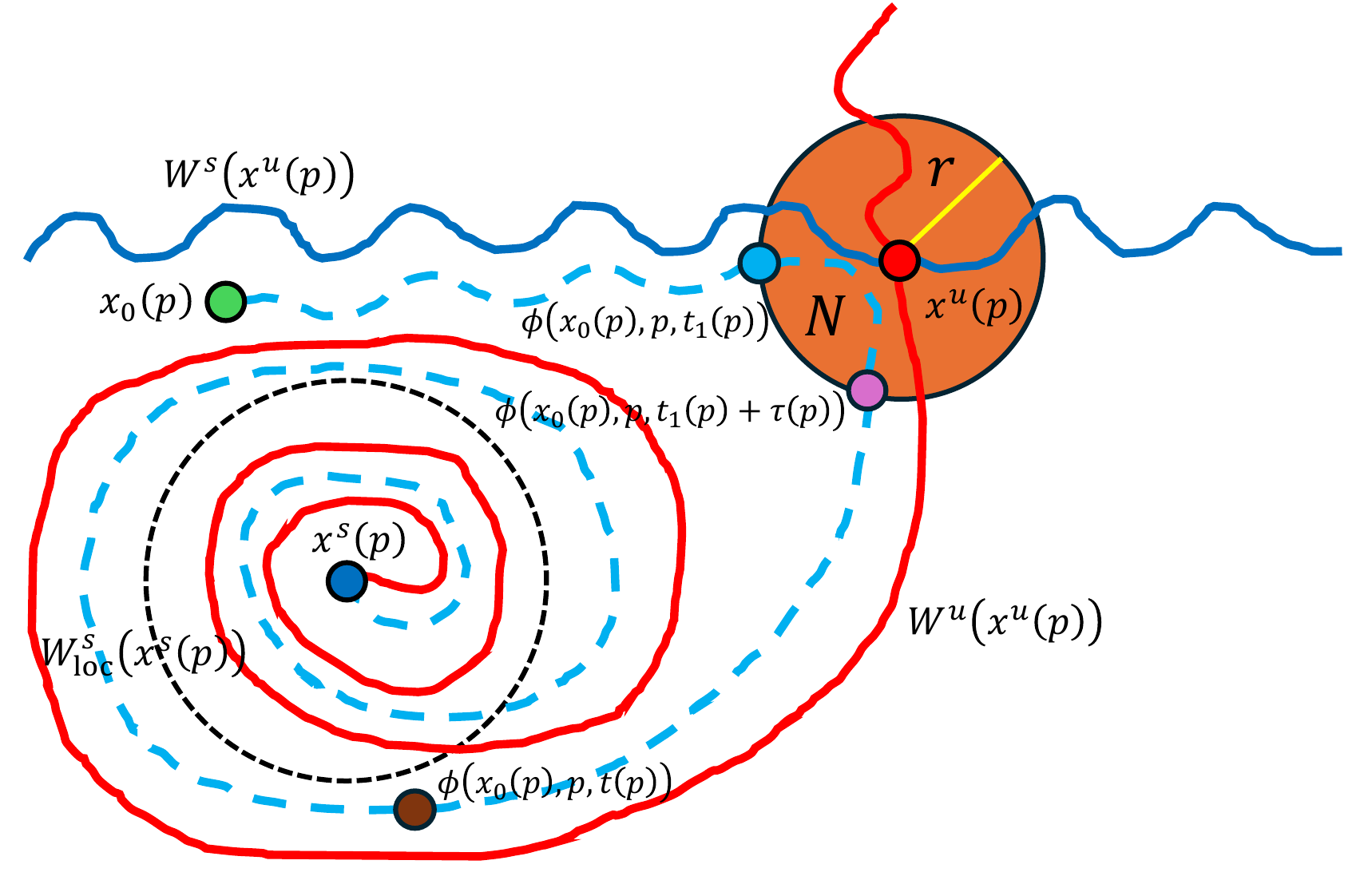}
\caption{The system trajectory (cyan dashed line) for parameter values near the recovery boundary from the IC $x_0\left(p\right)$ (green circle) and converges to the SEP $x^s\left(p\right)$ (dark blue circle). The stable manifold $W^s\left(x^u\left(p\right)\right)$ (blue solid line) and the unstable manifold of $W^u\left(x^u\left(p\right)\right)$ (red solid line) of the CUEP $x^u\left(p\right)$ (red circle), are shown.}
\label{fig:13}
\end{figure}

Let $\epsilon > 0$. To show that $\lim_{p \to p^*} F\left(p\right) = A\left(p^*\right)$, it suffices
  to show that there exists $\delta > 0$ such that $\left\|p-p^*\right\| < \delta$
  with $p \in R$ implies that $\left\|F(p)-A(p^*)\right\| < \epsilon$.
Let $p^*$ be the unique closest boundary parameter value to $p_0$ as in
Proposition~\ref{pro:2}, and let $x^u\left(p^*\right)$ be the CUEP.
Then, by \cite[Corollary 4.28]{fisher2022hausdorff}, $\lim_{t \to \infty} \phi\left(x_0\left(p^*\right),p^*,t\right) = x^u\left(p^*\right)$.
Since $f$ is $C^1$ its first derivatives are continuous, so this implies that 
$\lim_{t\rightarrow\infty}\frac{\partial f}{\partial x}\left(\phi\left(x_{0}\left(p^{*}\right),p^*,t\right)\right)= \frac{\partial f}{\partial x}\left(x^u\left(p^*\right)\right)= A\left(p^{*}\right)$.
As $f$ is $C^1$ at $\left(x^u\left(p^*\right),p^*\right)$, there exist $\delta>0$ and $r>0$ such
that whenever $\left\|p-p^{*}\right\|<\delta$ and
$\left\|x-x^{u}\left(p^{*}\right)\right\|<r$,
\begin{equation}
\label{eq:18}
\left\|\frac{\partial f}{\partial x}\left(x,p\right)-A\left(p^{*}\right)\right\|<\frac{\epsilon}{3}.
\end{equation}\par
Let $N=B_{r}\left(x^{u}\left(p^{*}\right)\right)$ denote
the closed ball of radius $r$ centered at $x^{u}\left(p^{*}\right)$.
Consider the system trajectory for $p = p^{*}$.
As $\lim_{t \to \infty} \phi\left(x_0\left(p^*\right),p^*,t\right) = x^u\left(p^*\right) \subset N$,
there exists some finite time at which
$\phi(x_0(p^*),p^*,t) \in \partial N$.
Shrink $r$ if necessary such that for $p = p^*$ the
system trajectory intersects $\partial N$ transversely at the final time
$\hat{t}_1 > 0$ at which it passes inwards through $\partial N$,
and such that $\partial N$ is
transverse to the local unstable manifold $W^u_{\text{loc}}\left(x^u\left(p^*\right)\right)$ with a single
point of intersection inside $W^s\left(x^s\left(p^*\right)\right)$.

Define the $C^1$ function
\begin{align*}
  &\Phi\left(p,t\right):=
  \\&\left[\phi\left(x_{0}\left(p\right),p,t\right)-x^{u}\left(p\right)\right]^{\intercal}
\left[\phi\left(x_{0}\left(p\right),p,t\right)-x^{u}\left(p\right)\right]-r^{2}.
\end{align*}
As $\phi\left(x_0\left(p^*\right),p^*,\hat{t}_1\right) \in \partial N=\partial B_r\left(x^u\left(p^*\right)\right)$,
$\Phi\left(p^*,\hat{t}_1\right)=0$.
Furthermore, $\frac{\partial \Phi}{\partial t}\left(p^{*},\hat{t}_{1}\right)
\neq 0$ (i.e., is full rank) because of the transversal intersection
between the system trajectory at $p = p^*$ and $\partial N$.
By the implicit function
theorem \cite[Theorem C.40]{lee2013smooth}, there exist open
neighborhoods $V_{0}\subseteq\mathbb{R}^{m}$ of $p^{*}$,
$W_{0}\subseteq\mathbb{R}$ of $\hat{t}_{1}$, and a $C^1$ function
$t_{1}: V_{0}\rightarrow W_{0}$ such that
$\Phi\left(p,t\right)=0$ for
$\left(p,t\right)\in V_{0}\times W_{0}$ if and only if
$t=t_{1}\left(p\right)$. This gives
$\hat{t}_{1}=t_{1}\left(p^{*}\right)$ so that
$\lim_{p\rightarrow p^{*}}t_{1}\left(p\right)=t_{1}\left(p^{*}\right)
=\hat{t}_{1}$.
Shrink
$\delta$ if necessary such that whenever
$\left\|p-p^{*}\right\|<\delta$,
\begin{equation}
\label{eq:19}
\left|t_{1}\left(p\right)-\hat{t}_{1}\right|<\epsilon.
\end{equation}

For $p \in R$ with $\left\|p-p^*\right\| < \delta$, define $\tau\left(p\right)$ such that $t_1\left(p\right) + \tau\left(p\right)$ is the final time at which the system trajectory passes outwards through $\partial N$. Note that this is finite and well-defined since $\lim_{t \to \infty} \phi\left(x_0\left(p\right),p,t\right) = x^s\left(p\right) \not\in \overline{N}$ and there exists at least one inward crossing of $\partial N$ at $t_1\left(p\right)$. Let $x_2\left(p\right)=\phi\left(x_0\left(p\right),p,t_1\left(p\right)+\tau\left(p\right)\right)$. By \cite[Theorem 4.32]{fisher2022hausdorff}, $\lim_{p\rightarrow p^{*}} \tau\left(p\right)=\infty$.


Let $\hat{x}_{2}$ be the single point of transversal intersection between $W_{\text{loc}}^{u}\left(x^{u}\left(p^{*}\right)\right)$ and $\partial N$ in $W^s\left(x^s\left(p^*\right)\right)$. We next show that $\lim_{p\rightarrow p^{*}}x_{2}\left(p\right)=\hat{x}_{2}$. So, let $\tilde{\epsilon} > 0$.
As $W_{\text{loc}}^u\left(x^u\left(p^*\right)\right)$ and $\partial N$ are compact and have a single point of transversal intersection in $W^s\left(x^s\left(p^*\right)\right)$, there exists $\hat{\epsilon} > 0$
such that for any manifold $M$ that satisfies
$d_{C_{1}}\left(M,W_{\text{loc}}^{u}\left(x^{u}\left(p^{*}\right)\right)\right)< \hat{\epsilon}$,
$M$ intersects $N$ transversely with a single point of intersection in
  $W^s\left(x^s\left(p^*\right)\right)$ and
$d_{C_{1}}\left(M\cap\partial N\cap W^s\left(x^s\left(p^*\right)\right),\hat{x}_{2}\right)<\tilde{\epsilon}$ \cite[Corollary A.3.18]{katok1999introduction}. Note that $M\cap\partial N \cap W^s\left(x^s\left(p^*\right)\right)$ is this single point of intersection.

 Consider $B_{\delta}\left(p^{*}\right)$, the open ball of radius $\delta$ centered at $p^{*}$. Let $\phi\left(x_{0}\left(B_{\delta}\left(p^{*}\right)\right),p,t\right):=\cup_{p\in B_{\delta}\left(p^{*}\right)}\phi\left(x_{0}\left(p\right),p,t\right)$,
 and note that $x_0\left(B_\delta\left(p^*\right)\right)$ is a $C^1$ manifold that is transverse to $
 W^s\left(x^u\left(p^*\right)\right)$ by Proposition \labelcref{pro:1}(ii)
Thus, by the the Inclination Lemma \cite{palis1969morse}, shrinking $\delta$ further if necessary,
there exists $T > 0$ such that
whenever $p \in B_{\delta}\left(p^*\right)$ and $t>T$,
\begin{equation}
\label{eq:20}
d_{C_{1}}\left(\phi\left(x_{0}\left(B_{\delta}\left(p^{*}\right)\right),p,t\right)\cap
N, W_{\text{loc}}^{u}\left(x^{u}\left(p\right)\right)\right)
<\frac{\hat{\epsilon}}{2},
\end{equation}
where $\phi\left(x_{0}\left(B_{\delta}\left(p^{*}\right)\right),p,t\right)\cap N$
refers to the connected component of this intersection that contains $\phi\left(x_{0}\left(p^{*}\right),p^{*},t\right)$.
Since $W_{\text{loc}}^{u}\left(x^{u}\left(p\right)\right)$ is
continuous in $p$, $\lim_{p\rightarrow p^{*}}W_{\text{loc}}^{u}\left(x^{u}\left(p\right)\right)=W_{\text{loc}}^{u}\left(x^{u}\left(p^{*}\right)\right)$. Shrink
$\delta$ if necessary such that whenever
$\left\|p-p^{*}\right\|<\delta$,
\begin{equation}
\label{eq:21}
d_{C_{1}}\left(W_{\text{loc}}^{u}\left(x^{u}\left(p\right)\right), W_{\text{loc}}^{u}\left(x^{u}\left(p^{*}\right)\right)\right)<\frac{\hat{\epsilon}}{2}.
\end{equation}
Thus, we have
\begin{align*}
&d_{C_{1}}\left(\phi\left(x_{0}\left(B_{\delta}\left(p^{*}\right)\right),p,t\right)\cap N,W_{\text{loc}}^{u}\left(x^{u}\left(p^{*}\right)\right)\right)\\
&\overset{\substack{\text{triangle}\\\text{inequality}}}{\le} d_{C_{1}}\left(\phi\left(x_{0}\left(B_{\delta}\left(p^{*}\right)\right),p,t\right)\cap N,W_{\text{loc}}^{u}\left(x^{u}\left(p\right)\right)\right)\\
&+d_{C_{1}}\left(W_{\text{loc}}^{u}\left(x^{u}\left(p\right)\right),W_{\text{loc}}^{u}\left(x^{u}\left(p^{*}\right)\right)\right)\\
&\overset{\substack{\text{\labelcref{eq:20}, \labelcref{eq:21}}}}{<}\frac{\hat{\epsilon}}{2}+\frac{\hat{\epsilon}}{2} = \hat{\epsilon}.
\end{align*}
By the above, this implies that for any $\left\|p-p^*\right\| < \delta$ and $t > T$, $\phi\left(x_0\left(B_\delta\left(p^*\right)\right),p,t\right)$ is transverse to $\partial N$ with a single point of intersection in $W^s\left(x^s\left(p^*\right)\right)$, and that $d_{C^1}\left(\phi\left(x_0\left(B_\delta\left(p^*\right)\right),p,t\right) \cap \partial N \cap W^s\left(x^s\left(p^*\right)\right), \hat{x}_2\right) < \tilde{\epsilon}$.
Since
$\lim_{p\rightarrow p^{*}}t_{1}\left(p\right)+\tau\left(p\right)=\infty$, shrink
$\delta$ if necessary such that whenever
$\left\|p-p^{*}\right\|<\delta$,
$t_{1}\left(p\right)+\tau\left(p\right)>T$.
As $x_2\left(p\right) = \phi\left(x_0\left(p\right),p,t_1\left(p\right)+\tau\left(p\right)\right) \in \partial N$ and $\phi\left(x_0\left(B_\delta\left(p^*\right)\right),p,t_1\left(p\right)+\tau\left(p\right)\right)$ has a single point of intersection with $\partial N$ in $W^s\left(x^s\left(p^*\right)\right)$, it must hold that $x_2\left(p\right)$ is this single point of intersection.
Thus, we have
\begin{align*}
&d_{C_{1}}\left(x_{2}\left(p\right),\hat{x}_{2}\right)=\\
&d_{C_{1}}\left(\phi\left(x_{0}\left(B_{\delta}\left(p^{*}\right)\right),p,t_{1}\left(p\right)+\tau\left(p\right)\right)\cap\partial N \cap W^s\left(x^s\left(p^*\right)\right)\right.\\
&\left.,\hat{x}_{2}\right)<\tilde{\epsilon}.
\end{align*}
Since $\tilde{\epsilon} > 0$ was arbitrary, this proves $\lim_{p\rightarrow p^{*}}x_{2}\left(p\right)=\hat{x}_{2}$.

Next, consider the system trajectory at time $t$ starting from the initial
condition $\hat{x}_{2}$, denoted by
$\phi\left(\hat{x}_{2},p^*,t\right)$.
As $\hat{x}_2 \in W^s\left(x^s\left(p^*\right)\right)$, there exists $\hat{t}_2 > 0$ such that
$\phi\left(\hat{x}_2,p^*,\hat{t}_2\right) \in \text{int } W^s_{\text{loc}}\left(x^s\left(p^*\right)\right)$.
As $W^s_{\text{loc}}\left(x^s\left(p\right)\right)$ varies continuously with $p$,
$\lim_{p \to p^*} x_2\left(p\right) = \hat{x}_2$, and the flow is continuous, shrinking
$\delta$ further if necessary implies that for $\left\|p-p^*\right\| < \delta$,
$\phi\left(x_2\left(p\right),p,\hat{t}_2\right) \in \text{int } W^s_{\text{loc}}\left(x^s\left(p\right)\right)$.
Recall that for all $\left\|p-p^*\right\| < \delta$, $W^s_{\text{loc}}\left(x^s\left(p\right)\right)$ is forward
invariant and
has $\frac{\partial f}{\partial x}\left(x\right)$ stable for all $x \in W^s_{\text{loc}}\left(x^s\left(p\right)\right)$.
Thus, for all $\left\|p-p^*\right\| < \delta$ and $t \geq \hat{t}_2$,
$\frac{\partial f}{\partial x}\left(\phi\left(x_2\left(p\right),p,t\right)\right)$ is stable.
For $p \in R$, as $t\left(p\right)$ is the supremum over time at which
$\frac{\partial f}{\partial x}\left(\phi\left(x_0\left(p\right),p,t\right)\right)$ is unstable, this implies that
\begin{equation}
  t\left(p\right) < t_1\left(p\right) + \tau\left(p\right) + \hat{t}_2
  \label{eq:22}
\end{equation}
for all $\left\|p-p^*\right\| < \delta$.
For $\left\|p-p^*\right\| < \delta$ with $p \in R$,
define the functions
\begin{align}
  \label{eq:23}
  \Psi\left(p\right) &:= 
\displaystyle\sup_{t\in\left[0,t_{1}\left(p\right)\right]}\left\|\frac{\partial f}{\partial x}\left(\phi\left(x_{0}\left(p\right),p,t\right)\right)-A\left(p^{*}\right)\right\|, 
\\
\label{eq:24}
\hat{\Psi}\left(p\right) &:= \displaystyle\sup_{t\in\left[0,\hat{t}_{2}\right]}\left\|\frac{\partial f}{\partial x}\left(\phi\left(x_{2}\left(p\right),p,t\right)\right)-A\left(p^{*}\right)\right\|.
\end{align}
As $f$ and the flow are $C^1$, $x_0\left(p\right)$ and $x_2\left(p\right)$ are continuous at $p^*$, the norm is continuous, so $\left\|\frac{\partial f}{\partial x}\left(\phi\left(x_{0}\left(p\right),p,t\right)\right)-A\left(p^{*}\right)\right\|$ and $\left\|\frac{\partial f}{\partial x}\left(\phi\left(x_{2}\left(p\right),p,t\right)\right)-A\left(p^{*}\right)\right\|$ are both continuous in $\left(p,t\right)$. Thus, as $\left[0,t_1\left(p\right)\right]$ and $\left[0,\hat{t}_2\right]$ are compact, nonempty, and continuous in $p$ (since $t_1\left(p\right)$ is continuous), by the maximum theorem \cite[Page 116]{berge1963topological},
$\Psi$ and $\hat{\Psi}$
are continuous at $p^*$. Thus,
\begin{align}
  \label{eq:25}
  \lim_{p \to p^*} \Psi\left(p\right) &=
\displaystyle\lim_{p\rightarrow p^{*}} \displaystyle\sup_{t\in\left[0,t_{1}\left(p\right)\right]}\left\|\frac{\partial f}{\partial x}\left(\phi\left(x_{0}\left(p\right),p,t\right)\right)-A\left(p^{*}\right)\right\|\nonumber\\
&=\displaystyle\sup_{t\in\left[0,\hat{t}_{1}\right]}\left\|\frac{\partial f}{\partial x}\left(\phi\left(x_{0}\left(p^{*}\right),p^*,t\right)\right)-A\left(p^{*}\right)\right\|\nonumber\\
&=: \Psi\left(p^*\right)
\end{align}
and
\begin{align}
  \label{eq:26}
  \lim_{p \to p^*} \hat{\Psi}(p) &=
\displaystyle\lim_{p\rightarrow p^{*}}\sup_{t\in\left[0,\hat{t}_{2}\right]}\left\|\frac{\partial f}{\partial x}\left(\phi\left(x_{2}\left(p\right),p,t\right)\right)-A\left(p^{*}\right)\right\|\nonumber\\
&=\displaystyle\sup_{t\in\left[0,\hat{t}_{2}\right]}\left\|\frac{\partial f}{\partial x}\left(\phi\left(\hat{x}_{2},p^{*},t\right)\right)-A\left(p^{*}\right)\right\|\nonumber\\
&=:\hat{\Psi}\left(p^*\right).
\end{align}

Since $t\left(p\right) > \tau\left(p\right) + t_1\left(p\right)$ and $\lim_{p \to p^*} \tau\left(p\right) = \infty$,
$\lim_{p\rightarrow p^{*}}t\left(p\right)=\infty$.
Thus, shrink $\delta$ further if necessary such that whenever
$\left\|p-p^{*}\right\|<\delta$, 
\begin{align}
  &t(p) > \frac{3\left(\hat{t}_1+\epsilon\right)\left(\Psi\left(p^*\right)+\epsilon\right)}{\epsilon}\label{eq:27},\\
  &t(p) > \frac{3 \hat{t}_2\left(\hat{\Psi}\left(p^*\right)+\epsilon\right)}{\epsilon}\label{eq:28},\\
  &\left|\Psi\left(p\right)-\Psi\left(p^*\right)\right| < \epsilon\label{eq:29},\\
  &\left|\hat{\Psi}\left(p\right)-\hat{\Psi}\left(p^*\right)\right| < \epsilon\label{eq:30},
\end{align}
where the final two equations follow from \labelcref{eq:25}-\labelcref{eq:26}.

By the definitions of $\Psi$ and $\hat{\Psi}$ in
\labelcref{eq:23} and \labelcref{eq:24},
we also have that for $t\in\left[0,t_{1}\left(p\right)\right]$,
$\left\|\frac{\partial f}{\partial x}\left(\phi\left(x_{0}\left(p\right),p,t\right)\right)-A\left(p^{*}\right)\right\| \leq \Psi\left(p\right)$,
and for $t\in\left[0,\hat{t}_2\right]$,
$\left\|\frac{\partial f}{\partial x}\left(\phi\left(x_{2}\left(p\right),p,t\right)\right)-A\left(p^{*}\right)\right\| \leq \hat{\Psi}\left(p\right)$.
Thus, for any $\left\|p-p^{*}\right\|<\delta$, we have
\begin{align}
&\left\|F\left(p\right)-A\left(p^*\right)\right\| \nonumber \\
&=\left\|\frac{1}{t\left(p\right)}\displaystyle\int_{0}^{t\left(p\right)}{\frac{\partial f}{\partial x}\left(\phi\left(x_{0}\left(p\right),p,t\right)\right)\,dt}-A\left(p^{*}\right)\right\| \nonumber \\ &=\left\|\frac{1}{t\left(p\right)}\displaystyle\int_{0}^{t\left(p\right)}{\left[\frac{\partial f}{\partial x}\left(\phi\left(x_{0}\left(p\right),p,t\right)\right)-A\left(p^{*}\right)\right]\,dt}\right\|\nonumber\\
&\leq\frac{1}{t\left(p\right)}\displaystyle\int_{0}^{t\left(p\right)}{\left\|\frac{\partial f}{\partial x}\left(\phi\left(x_{0}\left(p\right),p,t\right)\right)-A\left(p^{*}\right)\right\|\,dt}\nonumber\\ &=\frac{1}{t\left(p\right)}\left[\displaystyle\int_{0}^{t_{1}\left(p\right)}{\left\|\frac{\partial f}{\partial x}\left(\phi\left(x_{0}\left(p\right),p,t\right)\right)-A\left(p^{*}\right)\right\|\,dt}\right.\nonumber\\ &+\displaystyle\int_{t_{1}\left(p\right)}^{t_{1}\left(p\right)+\tau\left(p\right)}{\left\|\frac{\partial f}{\partial x}\left(\phi\left(x_{0}\left(p\right),p,t\right)\right)-A\left(p^{*}\right)\right\|\,dt}\nonumber\\ &+\left.\displaystyle\int_{t_{1}\left(p\right)+\tau\left(p\right)}^{t\left(p\right)}{\left\|\frac{\partial f}{\partial x}\left(\phi\left(x_{0}\left(p\right),p,t\right)\right)-A\left(p^{*}\right)\right\|\,dt}\right]\nonumber\\
&\overset{\labelcref{eq:22}}{<}
\frac{1}{t\left(p\right)}\left[\displaystyle\int_{0}^{t_{1}\left(p\right)}{\left\|\frac{\partial f}{\partial x}\left(\phi\left(x_{0}\left(p\right),p,t\right)\right)-A\left(p^{*}\right)\right\|\,dt}\right.\nonumber\\ &+\displaystyle\int_{t_{1}\left(p\right)}^{t_{1}\left(p\right)+\tau\left(p\right)}{\left\|\frac{\partial f}{\partial x}\left(\phi\left(x_{0}\left(p\right),p,t\right)\right)-A\left(p^{*}\right)\right\|\,dt}\nonumber\\ &+\left.\displaystyle\int_{t_{1}\left(p\right)+\tau\left(p\right)}^{t_1\left(p\right)+\tau\left(p\right)+\hat{t}_2}{\left\|\frac{\partial f}{\partial x}\left(\phi\left(x_{0}\left(p\right),p,t\right)\right)-A\left(p^{*}\right)\right\|\,dt}\right]\nonumber\\
&=
\frac{1}{t\left(p\right)}\left[\displaystyle\int_{0}^{t_{1}\left(p\right)}{\left\|\frac{\partial f}{\partial x}\left(\phi\left(x_{0}\left(p\right),p,t\right)\right)-A\left(p^{*}\right)\right\|\,dt}\right.\nonumber\\ &+\displaystyle\int_{t_{1}\left(p\right)}^{t_{1}\left(p\right)+\tau\left(p\right)}{\left\|\frac{\partial f}{\partial x}\left(\phi\left(x_{0}\left(p\right),p,t\right)\right)-A\left(p^{*}\right)\right\|\,dt}\nonumber\\ &+\left.\displaystyle\int_{0}^{\hat{t}_2}{\left\|\frac{\partial f}{\partial x}\left(\phi\left(x_{2}\left(p\right),p,t\right)\right)-A\left(p^{*}\right)\right\|\,dt}\right]\nonumber\\
&\overset{\labelcref{eq:18},\labelcref{eq:23},\labelcref{eq:24}}{<}
\frac{1}{t\left(p\right)} \left[\displaystyle\int_{0}^{t_{1}\left(p\right)}\Psi\left(p\right)\,dt+\displaystyle\int_{t_{1}\left(p\right)}^{t_{1}\left(p\right)+\tau\left(p\right)}{\frac{\epsilon}{3}\,dt}\right.\nonumber\\ &\left.+\displaystyle\int_{0}^{\hat{t}_{2}} \hat{\Psi}\left(p\right) \,dt \right]\nonumber\\
&= \f{t_1\left(p\right)\Psi\left(p\right)}{t\left(p\right)} + \f{\tau\left(p\right)}{t\left(p\right)}\f{\epsilon}{3}
+ \f{\hat{t}_2\hat{\Psi}\left(p\right)}{t\left(p\right)} \nonumber \\
&\overset{\labelcref{eq:19},\labelcref{eq:29},\labelcref{eq:30}}{<}
\f{\left(\hat{t}_1+\epsilon\right)\left(\Psi\left(p^*\right)+\epsilon\right)}{t\left(p\right)}
+ \f{\epsilon}{3}
+ \f{\hat{t}_2\left(\hat{\Psi}\left(p^*\right)+\epsilon\right)}{t\left(p\right)} \nonumber \\
&\overset{\labelcref{eq:27},\labelcref{eq:28}}{<}\frac{\left(\hat{t}_{1}+\epsilon\right)\left(\Psi\left(p^*\right)+\epsilon\right)}{\frac{3\left(\hat{t}_{1}+\epsilon\right)\left(\Psi\left(p^*\right)+\epsilon\right)}{\epsilon}}+\frac{\epsilon}{3}+\frac{\hat{t}_{2}\left(\hat{\Psi}\left(p^*\right)+\epsilon\right)}{\frac{3\hat{t}_{2}\left(\hat{\Psi}\left(p^*\right)+\epsilon\right)}{\epsilon}}\nonumber\\
&=\f{\epsilon}{3}
+ \f{\epsilon}{3} + \f{\epsilon}{3} = \epsilon,
\nonumber
\end{align}
so $\left\|F\left(p\right)-A\left(p^*\right)\right\| < \epsilon$ for all $\left\|p-p^*\right\| < \delta$ with $p \in R$,
which completes the proof of Theorem \labelcref{the:1}.

\end{proof}

\begin{proof}[Proof of Theorem \labelcref{the:2}]
  Recall that
$\lambda\left(p^{*}\right)$ is the simple unstable eigenvalue of $A\left(p^{*}\right)$ and $v\left(p^{*}\right)$ is its corresponding eigenvector.
Furthermore, $A\left(p^*\right)$ is hyperbolic by Proposition~\ref{pro:1} and its proof, so all eigenvalues of $A\left(p^*\right)$ other than $\lambda\left(p^*\right)$ are stable (i.e., no eigenvalues of $A\left(p^*\right)$ have zero real part). Since the eigenvalues of a matrix are continuous functions of the elements of that matrix \cite[Theorem 3.1.2]{ortega1990numerical}, and by Theorem~\ref{the:1} $\lim_{p \to p^*} F\left(p\right) = A\left(p^*\right)$, this implies that the eigenvalues of $F\left(p\right)$ converge to the eigenvalues of $A\left(p^*\right)$ in the limit as $p \to p^*$. Thus, there exists $\delta > 0$ sufficiently small such that $\left\|p-p^*\right\| < \delta$ implies that $F\left(p\right)$ is hyperbolic and has exactly one unstable eigenvalue, call it $\lambda\left(p\right)$. Let $v\left(p\right)$ denote the eigenvector associated with $\lambda\left(p\right)$. As the eigenvalues of $F\left(p\right)$ converge to those of $A\left(p^*\right)$ in the limit as $p \to p^*$, $\lambda\left(p\right)$ is the unique unstable eigenvalue of $F\left(p\right)$, and $\lambda\left(p^*\right)$ is the unique unstable eigenvalue of $A\left(p^*\right)$, this implies that $\lim_{p \to p^*} \lambda\left(p\right) = \lambda\left(p^*\right)$. Since $\lambda\left(p\right)$ is a simple (i.e., multiplicity one) eigenvalue of $F\left(p\right)$ and $v\left(p\right)$ is its associated eigenvector, and since $\lim_{p \to p^*} F\left(p\right) = A\left(p^*\right)$, by \cite[Theorem 3.1.3]{ortega1990numerical} $\lim_{p \to p^*} v\left(p\right) = v\left(p^*\right)$. This completes the proof of Theorem \labelcref{the:2}.
\end{proof}

\begin{proof}[Proof of Theorem \labelcref{the:3}]
The goal of this proof is to follow a similar argument as in the proof of Theorem~\ref{the:1}, which involves splitting the system trajectory for $p \in R$ with $p$ close to $p^*$ into three segments as illustrated in Fig. \labelcref{fig:14}. Recall the definition of $j\left(p\right)$ from \labelcref{eq:9}. We begin by showing that $\lim_{p \to p^*} j\left(p\right) = \infty$. Let $K>0$. As $\lim_{n \to \infty} T^n\left(x_0\left(p^*\right),p^*,h\right) = x^u\left(p^*\right)$, $\frac{\partial f}{\partial x}\left(x^u\left(p^*\right)\right) = A\left(p^*\right)$ is unstable and hyperbolic, and $\frac{\partial f}{\partial x}$ is continuous, there exists $i \in \mathbb{N}$ such that
$ih>K$ and $\frac{\partial f}{\partial x}\left(T^{i}\left(x_{0}\left(p^{*}\right),p^*,h\right)\right)$ is unstable. Since $\frac{\partial f}{\partial x}\left(T^{i}\left(x_{0}\left(p\right),p,h\right)\right)$ is continuous in $p$, there exists $\delta>0$ such that whenever $\left\|p-p^{*}\right\|<\delta$, $\frac{\partial f}{\partial x} \left(T^{i}\left(x_{0}\left(p\right),p,h\right)\right)$ is unstable. Hence, by the definition of $j\left(p\right)$ in \labelcref{eq:9}, $j\left(p\right) \geq ih > K$.  As $K$ was arbitrary, this implies that $\lim_{p\rightarrow p^{*}}j\left(p\right)=\infty$.

\begin{figure}[ht]
\centering   \includegraphics[width=\columnwidth]{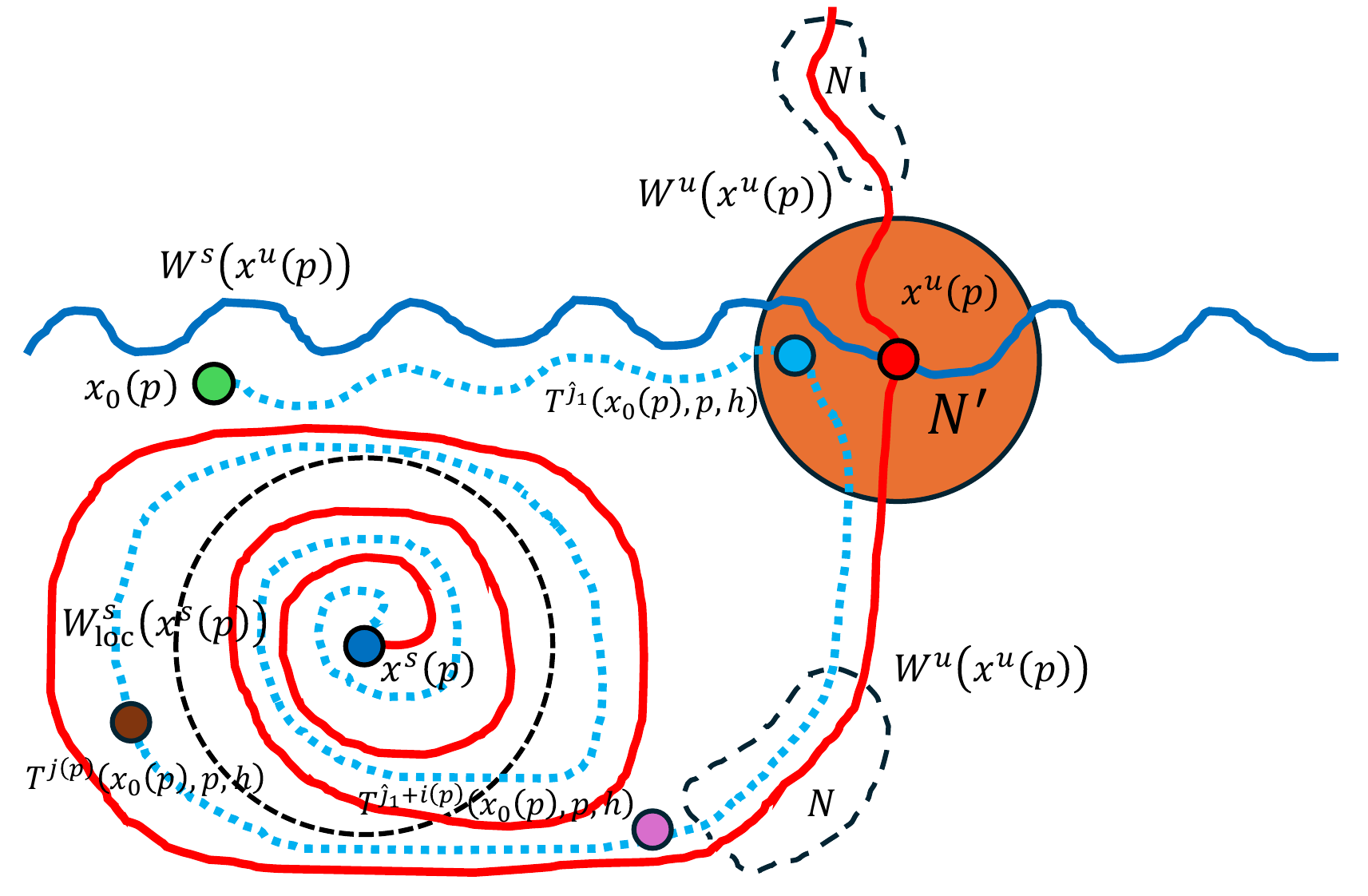}
\caption{The system trajectory in discrete time (cyan dotted line) for parameter values near the recovery boundary. This figure is similar to Fig. \labelcref{fig:13}, except we are considering the state obtained from the continuous time system \labelcref{eq:1}-\labelcref{eq:2} with new labels and definitions.}
\label{fig:14}
\end{figure}

As in the proof of Theorem~\ref{the:1}, there exist $r>0$ and $\delta > 0$ such that whenever $\left\|p-p^*\right\| < \delta$ and $\left\|x-x^u\left(p^*\right)\right\| < r$,
\begin{equation}
  \left\|\frac{\partial f}{\partial x}\left(x,p\right)-A\left(p^{*}\right)\right\|<\frac{\epsilon}{3}\label{eq:31}.
\end{equation}
Shrink $r$ further so that \labelcref{eq:31} holds for some $r' > r$. As in the proof of Theorem~\ref{the:1}, shrink $r$ further if necessary such that $\partial B_r\left(x^u\left(p^*\right)\right)$ is transverse to the local unstable manifold $W^u_{\text{loc}}\left(x^u\left(p^*\right)\right)$ with a single point of intersection $\hat{x}_2$ inside $W^s\left(x^s\left(p^*\right)\right)$ and a single point of intersection $\tilde{x}_2$ outside $W^s\left(x^s\left(p^*\right)\right)$. Let $\hat{x}_1 = T^{-1}\left(\hat{x}_2,p^*,h\right)$ and $\tilde{x}_1 = T^{-1}\left(\tilde{x}_2,p^*,h\right)$. Since $W^u\left(x^u\left(p^*\right)\right)$ is one-dimensional, let $\hat{D}$ be the subset of $W^u\left(x^u\left(p^*\right)\right)$ from $\hat{x}_1$ to $\hat{x}_2$, $\tilde{D}$ be the subset of $W^u\left(x^u\left(p^*\right)\right)$ from $\tilde{x}_1$ to $\tilde{x}_2$, and $D = \hat{D} \cup \tilde{D}$.
Then by \cite[Lemma 5.4]{fisher2023stability} (or, similarly, from the proof of \cite[Theorem 1.9]{palis1969morse}), shrinking $\delta$ further if necessary implies that there exist open neighborhoods $N$ and $N'$ such that $D \subset N$, $\overline{N}$ is disjoint from $W^s\left(x^u\left(p^*\right)\right)$ and $\partial W^s\left(x^s\left(p^*\right)\right)$, $\overline{N} \subset B_{r'}\left(x^u\left(p^*\right)\right)$, $\overline{N}$ is compact, and for $\left\|p-p^*\right\| < \delta$, $N'$ contains $x^u\left(p\right)$ and $\bigcup_{i = 0}^{\infty} T^{-i}\left(N,p,h\right)\cup W^s\left(x^u\left(p\right)\right)$ contains $N'$ \cite[Lemma 5.1]{fisher2022hausdorff}.
Let $\hat{N} = \left(N'\cup N\right) \cap B_{r'}\left(x^u\left(p^*\right)\right)$, and note that
$N \subset \hat{N}$.

Let $\hat{j}_{1}$ be the final time index at which $T^n\left(x_{0}\left(p^{*}\right),p^*,h\right)$ enters the interior of $\hat{N}$. Since $T^{\hat{j}_1}\left(x_{0}\left(p\right),p,h\right)$ is continuous in $p$, shrink $\delta$ if necessary such that whenever $\left\|p-p^{*}\right\|<\delta$, $T^{\hat{j}_{1}}\left(x_{0}\left(p\right),p,h\right)$ enters the interior of $\hat{N}$.
For $p \in R$ with $\left\|p-p^*\right\| < \delta$, define $i\left(p\right)$ 
  such that $\hat{j}_1 + i\left(p\right)$ is the final time index at which
  $T^n\left(x_0\left(p\right),p,h\right)$ exits $N$, and let $x_2\left(p\right) = T^{\hat{j}_1+i\left(p\right)}\left(x_0\left(p\right),p,h\right)$.
  Note that $i\left(p\right)$ is well-defined since for $p \in R$ with
  $\left\|p-p^*\right\| < \delta$, $T^{\hat{j}_1}\left(x_0\left(p\right),p,h\right) \in \hat{N}$, so since
  $T^{\hat{j}_1}\left(x_0\left(p\right),p,h\right)$ is not in $W^s\left(x^u\left(p\right)\right)$, it must be in
  $T^{-i}\left(N,p,h\right)$ 
  for some positive integer
  $i$.
  This implies that $T^{\hat{j}_1+n}\left(x_0\left(p\right),p,h\right)$ must pass through $N$ after a
  finite number of time steps, before ultimately converging to $x^s\left(p\right)$.
  Let $S = \overline{N} \cap \overline{W}^s\left(x^s\left(p^*\right)\right)$,
  which is compact.
  Then $S \subset W^s(x^s(p^*))$ since $\overline{N}$ and
  $\partial W^s\left(x^s\left(p^*\right)\right)$ are disjoint.
  Then as $S$ is compact and $T^n$ is continuous, there exists a time index
  $\hat{j}_2$ sufficiently large such that
  $T^{\hat{j}_2}\left(S,p^*,h\right) \subset \text{int } W^s_{\text{loc}}\left(x^s\left(p^*\right)\right)$.
  As $W^s_{\text{loc}}\left(x^s\left(p\right)\right)$ is continuous in $p$, $S$ is compact, and $T^{\hat{j}_2}$
  is continuous, shrink $\delta$ further if necessary such that
  $\left\|p-p^*\right\| < \delta$ with $p \in R$ implies that
  $T^{\hat{j}_2}\left(S,p,h\right) \subset \text{int } W^s_{\text{loc}}\left(x^s\left(p\right)\right)$.
  As $\text{int } W^s_{\text{loc}}\left(x^s\left(p\right)\right)$ is forward invariant and the Jacobian
  is stable for each $x \in W^s_{\text{loc}}\left(x^s\left(p\right)\right)$, this implies that
  for $p \in R$ with $\left\|p-p^*\right\| < \delta$,
  \begin{equation}
    j\left(p\right) < \hat{j}_1 + i\left(p\right) + \hat{j}_2\label{eq:32}.
  \end{equation}
  For $p \in R$ with $\left\|p-p^*\right\| < \delta$, define the functions
  \begin{align}
    \Psi\left(p\right) &:= \max_{n \in \left\{0, \hdots ,\hat{j}_1-1\right\}}
    \left\|\frac{\partial f}{\partial x}\left(T^{n}\left(x_{0}\left(p\right),p,h\right)\right)
    -A\left(p^{*}\right)\right\|\label{eq:33},
    \\
    \hat{\Psi}\left(p\right) &:= \max_{n \in \left\{0, \hdots ,\hat{j}_2\right\}}
    \left\|\frac{\partial f}{\partial x}\left(T^{n}\left(x_{2}\left(p\right),p,h\right)\right)
    -A\left(p^{*}\right)\right\|\label{eq:34}.
  \end{align}
  Note that $\Psi\left(p\right)$ and $\hat{\Psi}\left(p\right)$, as the maximums of a finite
  number of continuous functions, are themselves continuous.
  We have
  \begin{align*}
    &\lim_{p \to p^*} \Psi\left(p\right)=\Psi\left(p^*\right)\\
    &:=
    \max_{n \in \left\{0, \hdots ,\hat{j}_1-1\right\}}
    \left\|\frac{\partial f}{\partial x}\left(T^{n}\left(x_{0}\left(p^*\right),p^*,h
    \right)\right)-A\left(p^{*}\right)\right\|.
  \end{align*}
  As $\max_{n \in \left\{0, \hdots ,\hat{j}_2\right\}}
  \left\|\frac{\partial f}{\partial x}\left(T^{n}\left(x_{2}\left(p\right),p,h\right)\right)-A\left(p^{*}\right)\right\|$ is continuous, $x_2\left(p\right) \in N$, and $\overline{N}$ is compact,
  \begin{align}
    \hat{\Psi}\left(p\right) &\leq
    \sup_{x \in \overline{N}} \max_{n \in \left\{0, \hdots ,\hat{j}_2\right\}}
    \left\|\frac{\partial f}{\partial x}\left(T^{n}\left(x,p,h\right)\right)
    -A\left(p^{*}\right)\right\| \nonumber \\ &=:
    \hat{\Psi}\left(p^*\right) < \infty \label{eq:35}.
  \end{align}

Since $\lim_{p \to p^*} j\left(p\right) = \infty$,
shrink $\delta$ if necessary such that whenever $p \in R$ with
$\left\|p-p^{*}\right\|<\delta$,
\begin{align}
  & j\left(p\right) > \f{3\hat{j}_1\left(\Psi\left(p^*\right)+\epsilon\right)}{\epsilon}\label{eq:36},\\
  & j\left(p\right) > \f{3\left(\hat{j}_2+1\right) \hat{\Psi}\left(p^*\right)}{\epsilon} \label{eq:37}, \\
  &\left|\Psi\left(p\right)-\Psi\left(p^*\right)\right| < \epsilon
  \label{eq:38}.
\end{align}
For $n \in \left\{\hat{j}_1, \hdots, \hat{j}_1 + i\left(p\right)-1\right\}$,
since $T^n\left(x_0\left(p\right),p,h\right) \in B_{r'}\left(x^u\left(p^*\right)\right)$, by \labelcref{eq:31} we have
\begin{equation}
  \left\|\frac{\partial f}{\partial x}\left(T^n\left(x_0\left(p\right),p,h\right)\right)-A\left(p^*\right)\right\| <
  \f{\epsilon}{3}
  \label{eq:39}.
\end{equation}

Thus, we have whenever $\left\|p-p^{*}\right\|<\delta$,
\begin{align*}
  &\left\|\hat{F}\left(p\right)-A\left(p^*\right)\right\|\\
  &=\left\|\frac{1}{j\left(p\right)}\displaystyle\sum_{n=0}^{j\left(p\right)}
  \frac{\partial f}{\partial x}\left(T^{n}\left(x_{0}\left(p\right),p,h\right)\right)-A\left(p^{*}\right)\right\|\nonumber\\
&=\left\|\frac{1}{j\left(p\right)}\displaystyle\sum_{n=0}^{j\left(p\right)}\left[
    \frac{\partial f}{\partial x}\left(T^{n}\left(x_{0}\left(p\right),p,h\right)\right)-A\left(p^{*}\right)\right]\right\|\nonumber\\
 &\leq \frac{1}{j\left(p\right)} \sum_{n=0}^{j\left(p\right)}
  \left\|\displaystyle
  \frac{\partial f}{\partial x}\left(T^{n}\left(x_{0}\left(p\right),p,h\right)\right)-A\left(p^{*}\right)\right\|\nonumber\\
 &= \frac{1}{j\left(p\right)} \left[\sum_{n=0}^{\hat{j}_1-1}
  \left\|\displaystyle
  \frac{\partial f}{\partial x}\left(T^{n}\left(x_{0}\left(p\right),p,h\right)\right)-A\left(p^{*}\right)\right\| \right. \nonumber \\
&+\sum_{n=\hat{j}_1}^{\hat{j}_1+i\left(p\right)-1}
  \left\|\displaystyle
  \frac{\partial f}{\partial x}\left(T^{n}\left(x_{0}\left(p\right),p,h\right)\right)-A\left(p^{*}\right)\right\| \nonumber \\
&\left.+\sum_{n=\hat{j}_1+i\left(p\right)}^{j\left(p\right)}
  \left\|\displaystyle
  \frac{\partial f}{\partial x}\left(T^{n}\left(x_{0}\left(p\right),p,h\right)\right)-A\left(p^{*}\right)\right\| \right] \nonumber\\
  &\overset{\labelcref{eq:32}}{<}
  \frac{1}{j\left(p\right)} \left[\sum_{n=0}^{\hat{j}_1-1}
  \left\|\displaystyle
  \frac{\partial f}{\partial x}\left(T^{n}\left(x_{0}\left(p\right),p,h\right)\right)-A\left(p^{*}\right)\right\| \right. \nonumber \\
&+\sum_{n=\hat{j}_1}^{\hat{j}_1+i\left(p\right)-1}
  \left\|\displaystyle
  \frac{\partial f}{\partial x}\left(T^{n}\left(x_{0}\left(p\right),p,h\right)\right)-A\left(p^{*}\right)\right\| \nonumber \\
&\left.+\sum_{n=\hat{j}_1+i\left(p\right)}^{\hat{j}_1+i\left(p\right)+\hat{j}_2}
  \left\|\displaystyle
  \frac{\partial f}{\partial x}\left(T^{n}\left(x_{0}\left(p\right),p,h\right)\right)-A\left(p^{*}\right)\right\| \right] \nonumber\\
  &=
  \frac{1}{j\left(p\right)} \left[\sum_{n=0}^{\hat{j}_1-1}
  \left\|\displaystyle
  \frac{\partial f}{\partial x}\left(T^{n}\left(x_{0}\left(p\right),p,h\right)\right)-A\left(p^{*}\right)\right\| \right. \nonumber \\
&+\sum_{n=\hat{j}_1}^{\hat{j}_1+i\left(p\right)-1}
  \left\|\displaystyle
  \frac{\partial f}{\partial x}\left(T^{n}\left(x_{0}\left(p\right),p,h\right)\right)-A\left(p^{*}\right)\right\| \nonumber \\
&\left.+\sum_{n=0}^{\hat{j}_2}
  \left\|\displaystyle
  \frac{\partial f}{\partial x}\left(T^{n}\left(x_{2}\left(p\right),p,h\right)\right)-A\left(p^{*}\right)\right\| \right] \nonumber\\
  &\overset{\labelcref{eq:33},\labelcref{eq:34},\labelcref{eq:39}}{<}\frac{1}{j\left(p\right)} \left[\sum_{n=0}^{\hat{j}_1-1} \Psi\left(p\right)+\sum_{n=\hat{j}_1}^{\hat{j}_1+i\left(p\right)-1} \frac{\epsilon}{3}+\sum_{n=0}^{\hat{j}_2} \hat{\Psi}\left(p\right)\right]\nonumber\\
&\overset{\labelcref{eq:35},\labelcref{eq:38}}{<}
  \frac{1}{j\left(p\right)} \left[\sum_{n=0}^{\hat{j}_1-1} \left(\Psi\left(p^*\right)+\epsilon\right)
+\sum_{n=\hat{j}_1}^{\hat{j}_1+i\left(p\right)-1} \frac{\epsilon}{3}  \right.
 \nonumber \\
&\left.+\sum_{n=0}^{\hat{j}_2} \hat{\Psi}\left(p^*\right)
 \right] \nonumber\\
  &= \frac{\hat{j}_1\left(\Psi\left(p^*\right)+\epsilon\right)}{j\left(p\right)}
  + \frac{i\left(p\right)}{j\left(p\right)} \frac{\epsilon}{3}
  + \frac{\left(\hat{j}_2+1\right)\hat{\Psi}\left(p^*\right)}{j\left(p\right)} \nonumber \\
  &\overset{\labelcref{eq:36},\labelcref{eq:37}}{<}\frac{\hat{j}_1\left(\Psi\left(p^*\right)+\epsilon\right)}{\frac{3\hat{j}_1\left(\Psi\left(p^*\right)+\epsilon\right)}{\epsilon}}+\frac{\epsilon}{3}+\frac{\left(\hat{j}_2+1\right)\hat{\Psi}\left(p^*\right)}{\frac{3\left(\hat{j}_2+1\right)\hat{\Psi}\left(p^*\right)}{\epsilon}}\nonumber\\
  &=\frac{\epsilon}{3}
  + \frac{\epsilon}{3} + \frac{\epsilon}{3}=\epsilon\nonumber,
\end{align*}
so $\left\|\hat{F}\left(p\right)-A\left(p^*\right)\right\| < \epsilon$ for all $\left\|p-p^*\right\| < \delta$
with $p \in R$, which completes the proof of \labelcref{eq:11a}.
\labelcref{eq:11b} and \labelcref{eq:11c} can be
proved by following analogous reasoning as in the proof of Theorem
\labelcref{the:2}. This completes the proof of Theorem
\labelcref{the:3}.
\end{proof}

\begin{proof}[Proof of Theorem \labelcref{the:4}]
We first prove $\lim_{h\rightarrow 0}p^*\left(h\right)=p^*$.
Recall that $x_0\left(p^*\right) \in W^s\left(x^u\left(p^*\right)\right)$, where $x^u\left(p^*\right)$ is the CUEP.
Thus, there exists $T > 0$ such that
$x_0\left(p^*\right) \in \phi\left(W^s_{\text{loc}}\left(x^u\left(p^*\right)\right),p^*,-T\right) \subset
\phi\left(W^s_{\text{loc}}\left(x^u\left(J\right)\right),J,-T\right)$.
By Assumption~\ref{ass:1}, $W^s\left(x^u\left(J\right)\right)$ and $x_0\left(J\right)$ are transverse, which
implies that $\phi\left(W^s_{\text{loc}}\left(x^u\left(J\right)\right),J,-T\right)$ and $x_0\left(J\right)$ are
transverse, and that $x_0\left(p^*\right)$ is a point of their transversal intersection.
Then for any $\epsilon > 0$, by \cite[Proposition A.3.16]{katok1999introduction}
(i.e., the stability of transversal intersections) there exists a
$\hat{\delta} > 0$ such that for any manifold $M$ that satisfies
$d_{C^1}\left(M,\phi\left(W^s_{\text{loc}}\left(x^u\left(J\right)\right),J,-T\right)\right) < \hat{\delta}$,
there exists a point $\left(x,p\right) \in M \cap x_0\left(J\right)$ with
$\left\|\left(x,p\right)-\left(x_0\left(p^*\right),p^*\right)\right\| < \epsilon$.
As $\phi\left(\cdot,\cdot,-T\right)$ is $C^1$, there exists $\delta > 0$ such
that for any manifold $M$ that satisfies
$d_{C^1}\left(M,W^s_{\text{loc}}\left(x^u\left(J\right)\right)\right) < \delta$,
$d_{C^1}\left(\phi\left(M,-T\right),
\phi\left(W^s_{\text{loc}}\left(x^u\left(J\right)\right),-T\right)\right) < \hat{\delta}$,
so $\phi\left(M,-T\right)$ intersects $x_0\left(J\right)$ at a point within an
$\epsilon$ distance of $\left(x_0\left(p^*\right),p^*\right)$.

Recall that $T^n\left(x,p,h\right) = \phi\left(x,p,nh\right)$ for any state $x$, parameter value $p$,
and time step $n$.
We refer to $T$ as the exact discrete time map with time step $h$.
Let $T_d$ denote the approximation of $T$ that is obtained using numerical
integration.
Fix some time $\hat{t} > 0$.  For any $h > 0$, let $\hat{n}\left(h\right)$ be the minimum
integer $n$ such that $nh \geq \hat{t}$.
Over any compact set $K \times J' \subset \mathbb{R}^n \times J$,
define the maps $\hat{T}$ and $\hat{T}_d$ by
$\hat{T}\left(x,p,h\right) = T^{\hat{n}\left(h\right)}\left(x,p,h\right)$ and
$\hat{T}_d\left(x,p,h\right) = T^{\hat{n}\left(h\right)}_d\left(x,p,h\right)$, respectively,
for all $\left(x,p\right) \in K \times J'$ and $h > 0$.
Then, since the approximation error of the numerical integration approaches zero
as $h \to 0$, we have that 
$\lim_{h \to 0} d_{C^1}\left(\hat{T}_d\left(\cdot,\cdot,h\right),\hat{T}\left(\cdot,\cdot,h\right)\right) = 0$.
Furthermore, as $\lim_{h \to 0} h\hat{n}\left(h\right) = \tilde{t}$, and since $\hat{T}$
represents the exact discrete time map, we also have that
$\lim_{h \to 0} d_{C^1}\left(\hat{T}\left(\cdot,\cdot,h\right),\phi\left(\cdot,\cdot,\tilde{t}\right)\right) = 0$,
where $\phi\left(\cdot,\cdot,\tilde{t}\right)$ is the time-$\tilde{t}$ map for the flow
$\phi$, and is itself a discrete time map.
Thus, combining these implies that
\begin{align}
  \lim_{h \to 0} d_{C^1}\left(\hat{T}_d\left(\cdot,\cdot,h\right),\phi\left(\cdot,\cdot,\tilde{t}\right)\right) = 0
  \label{eq:limh}.
\end{align}

Note that $x^u\left(p\right)$ is a hyperbolic fixed point of $\phi\left(\cdot,\cdot,\tilde{t}\right)$
for all $p \in J$, and that $W^s_{\text{loc}}\left(x^u\left(J\right)\right)$ is its local stable
manifold
under $\phi\left(\cdot,\cdot,\tilde{t}\right)$, since $\phi\left(\cdot,\cdot,\tilde{t}\right)$
is the exact time-$\tilde{t}$ map for the flow $\phi$.
Thus, there exists $\tilde{\delta} > 0$ such that for any map $\tilde{T}$ with
$d_{C^1}\left(\tilde{T},\phi\left(\cdot,\cdot,\tilde{t}\right)\right) < \tilde{\delta}$,
$\tilde{T}$ possesses a hyperbolic fixed point
$\tilde{x}^u$ near $x^u$ and
$d_{C^1}\left(W^s_{\text{loc}}\left(\tilde{x}^u\left(J\right)\right),W^s_{\text{loc}}\left(x^u\left(J\right)\right)\right) < \delta$.
By \labelcref{eq:limh}, this implies that there exists $\hat{h} > 0$ such that for
any $h \in \left(0,\hat{h}\right)$, $\hat{T}_d\left(\cdot,\cdot,h\right)$ possesses a hyperbolic fixed
point $x_h^u$ near $x^u$
and $d_{C^1}\left(W^s_{\text{loc}}\left(x_h^u\left(J\right)\right),W^s_{\text{loc}}\left(x^u\left(J\right)\right)\right) < \delta$.
Thus, by the above there exists a point
$\left(x_h,p_h\right) \in W^s_{\text{loc}}\left(x_h^u\left(J\right)\right) \cap x_0\left(J\right)$ with
$\left\|\left(x_h,p_h\right)-\left(x_0\left(p^*\right),p^*\right)\right\| < \epsilon$.
As $\epsilon > 0$ was arbitrary, this implies that the points $\left(x_h,p_h\right)$
can be selected such that $\lim_{h \to 0} p_h = p^*$.
Furthermore,
as $x_0\left(p_h\right) = \left(x_h,p_h\right) \in W^s_{\text{loc}}\left(x_h^u\left(J\right)\right)$ for the map
$\hat{T}_d\left(\cdot,\cdot,h\right)$, $x_0\left(p_h\right) \not\in W^s_{\text{loc}}\left(x^s_h\left(J\right)\right)$ for
this map, where $x^s_h$ represents the SEP of $\hat{T}_d\left(\cdot,\cdot,h\right)$
near $x^s$, so $p_h \not\in R$ for this map.
Also, $x_h^u$ is a hyperbolic fixed point for both $\hat{T}_d\left(\cdot,\cdot,h\right)$
and $T_d\left(\cdot,\cdot,h\right)$, and $W^s_{\text{loc}}\left(x_h^u\left(J\right)\right)$ is its local stable
manifold under both maps, so $p_h \not\in R$ for the map
$T_d\left(\cdot,\cdot,h\right)$ as well.

Assume towards a contradiction that $\lim_{h\to 0}p^*\left(h\right) \neq p^*$.
Then there must exist a monotonically decreasing sequence
$\left\{h_n\right\}_{n=1}^\infty$ with
$\lim_{n\to\infty}h_n=0$ such that $\lim_{n \to \infty} p^*\left(h_n\right) \neq p^*$.
Since $p^*\left(h_n\right)$ is the unique closest parameter value in $\partial R$ to
$p_0$ for the map $T_d\left(\cdot,\cdot,h_n\right)$, and since $p_{h_n} \not\in R$
for the map $T_d\left(\cdot,\cdot,h_n\right)$, for each integer $n$ we have
\begin{align*}
  \left\|p^*\left(h_n\right)-p_0\right\| \leq \left\|p_{h_n}-p_0\right\|
  \leq \sup_{n \geq 1} \left\|p_{h_n}-p_0\right\| =: r < \infty,
\end{align*}
where the supremum is finite since $\lim_{n \to \infty} p_{h_n} = p^*$.
Therefore, the
sequence $\left\{p^*\left(h_n\right)\right\}_{n=1}^{\infty}$ is
contained in the closed ball $\overline{B}_r\left(p_0\right)$, which is compact.
Thus, $\left\{p^*\left(h_n\right)\right\}_{n=1}^{\infty}$ must have a convergent
subsequence.
If every convergent subsequence of
$\left\{p^*\left(h_n\right)\right\}_{n=1}^{\infty}$ converges to $p^*$,
then (since the sequence is contained in a compact set)
$\lim_{n \to \infty} p^*\left(h_n\right) = p^*$ \cite{sanfelice2014asymptotic}, and we obtain a contradiction,
so suppose there exists a convergent subsequence
$\left\{p^*\left(h_{n_m}\right)\right\}_{m=1}^\infty$ with
$\lim_{m \to \infty} p^*\left(h_{n_m}\right) = \hat{p} \neq p^*$.
Since
$\left\|p^*\left(h_{n_m}\right)-p_0\right\|\leq \left\|p_{h_{n_m}}-p_0\right\|$
for all $m$, taking the limit as $m \to \infty$ implies that
$\left\|\hat{p}-p_0\right\|\leq \left\|p^*-p_0\right\|$.
By Proposition~\ref{pro:2}, $p^*$ is the unique closest parameter value in
$\partial R$ to $p_0$ under the map $\phi\left(\cdot,\cdot,\tilde{t}\right)$.
Therefore, since $\hat{p} \neq p^*$ and is at least as close to $p_0$ as $p^*$,
we must have $\hat{p} \in R$ under the map $\phi\left(\cdot,\cdot,\tilde{t}\right)$.
Thus, $x_0\left(\hat{p}\right) \in W^s\left(x^s\left(\hat{p}\right)\right)$, so there exists $\hat{T} > 0$
such that
$x_0\left(\hat{p}\right) \in \text{int }
\phi\left(W^s_{\text{loc}}\left(x^s\left(\hat{p}\right)\right),\hat{p},-\hat{T}\right)$.

As $\phi$ and $x_0$ are $C^1$, and by \labelcref{eq:limh}, there exist $h' > 0$ and
$\delta' > 0$ such that $h \in \left(0,h'\right)$ and $\left\|p-\hat{p}\right\| < \delta'$
implies that $W^s_{\text{loc}}\left(x^s_h\left(p\right)\right)$ under $\hat{T}_d\left(\cdot,p,h\right)$ is
sufficiently $C^1$ close to $W^s_{\text{loc}}\left(x^s\left(\hat{p}\right)\right)$ under
$\phi\left(\cdot,\hat{p},\tilde{t}\right)$ such that 
$x_0\left(p\right) \in \phi\left(W^s_{\text{loc}}\left(x^s_h\left(p\right)\right),p,-\hat{T}\right)$.
Thus, for $m$ sufficiently large,
$x_0\left(p^*\left(h_{n_m}\right)\right) \in \phi\left(W^s_{\text{loc}}\left(x^s_{h_{n_m}}\left(p^*\left(h_{n_m}\right)\right)\right),p^*\left(h_{n_m}\right),
-\hat{T}\right)$.
Thus, for $m$ sufficiently large, $p^*\left(h_{n_m}\right) \in R$ under
$T_d\left(\cdot,\cdot,h_{n_m}\right)$, which contradicts
the definition of $p^*\left(h\right)$ as the closest point in $\partial R$ to $p_0$
under $T_d\left(\cdot,\cdot,h\right)$.
Therefore, we must have that $\lim_{h \to 0} p^*\left(h\right) = p^*$, which proves
\labelcref{eq:12a}.

Then, \labelcref{eq:12b} can be proved by following analogous reasoning as in the
proof of Theorem~\ref{the:3}, with the following changes:
\begin{itemize}
\item All instances of $T$ should be replaced with $T_d\left(\cdot,\cdot,h\right)$,
  including the definitions of $j\left(p\right)$ in \labelcref{eq:9} and $\hat{F}\left(p\right)$ in \labelcref{eq:10}.
\item All instances of $p^*$ should be replaced with
  $p^*\left(h\right)$.
\item For $h>0$ sufficiently small, there exists $x^s_h\left(p\right)$ near $x^s\left(p\right)$
  stable, and $x^u_h\left(p\right)$ near $x^u\left(p\right)$ unstable,
  which are
  hyperbolic fixed points of $T_d\left(\cdot,\cdot,h\right)$ such that
  $\lim_{n \to \infty} T^n_d\left(x_0\left(p^*\left(h\right)\right),p^*\left(h\right),h\right) = x^u_h\left(p^*\left(h\right)\right)$.
\item Since $W^u\left(x^u_h\left(p^*\left(h\right)\right)\right)$ is one-dimensional, define $\hat{D}$ and
$\tilde{D}$ to be the subsets of $W^u\left(x^u_h\left(p^*\left(h\right)\right)\right)$ from
$\hat{x}_1$ to $T_d\left(\hat{x}_1,p^*\left(h\right),h\right)$, and from
  $\tilde{x}_1$ to $T_d\left(\tilde{x}_1,p^*\left(h\right),h\right)$.
\end{itemize}

By continuity of $\frac{\partial f}{\partial x}$ and $x^u_h\left(p\right)$, and by
\labelcref{eq:12a} and \labelcref{eq:12b},
$\lim_{h\rightarrow0}\lim_{p\rightarrow p^*\left(h\right)}\tilde{F}\left(p\right)=\lim_{h\rightarrow0}\frac{\partial f}{\partial x}\left(x^u_h\left(p^*\left(h\right)\right)\right)=\frac{\partial f}{\partial x}\left(x^u\left(p^*\right)\right)=A\left(p^*\right)$,
and this proves \labelcref{eq:12c}. \labelcref{eq:13a} and \labelcref{eq:13b} can
be proved by following analogous reasoning as in the proof of Theorem
\labelcref{the:2}. This completes the proof of Theorem
\labelcref{the:4}.

\end{proof}

\section{Conclusion}\label{sec:con}


This work develops a computationally efficient method with rigorous
convergence guarantees for numerically computing the mode of
instability for parameterized nonlinear systems while avoiding the
challenge of identifying the CUEP. For parameter values where the
system recovers, it averages the Jacobian along the system trajectory
from the IC until the final time at which the Jacobian transitions
from unstable to stable.
It is shown that as boundary parameter values are approached
from within the recovery region, this average of the Jacobians
converges to the true Jacobian at the CUEP, and it has a unique
unstable eigenvalue whose corresponding eigenvector converges to the
mode of instability. Under the same conditions, the approximation of
the average of the Jacobians obtained from numerical integration, and
its eigenvector corresponding to its unique unstable eigenvalue, are shown to
converge to the true Jacobian at the CUEP and the mode of instability,
respectively, as the step size approaches zero.

The method was first validated on the
damped, driven, nonlinear pendulum example subject to a
disturbance. By directly finding the CUEP and analytically computing the
mode of instability for this low dimensional system, it was shown that the
mode of instability could be
accurately computed using the proposed algorithm in this case, thereby
validating the proposed approach. Then the algorithm was applied to compute the
mode of instability for the IEEE 9-bus power system subject to a
temporary short circuit. 
For
this higher dimensional power system model, it was no longer
straightforward to identify the CUEP and the mode of instability analytically.
The method was used to identify the mode of instability in high dimensional
parameter space for this system.
The identified modes of instability revealed non-intuitive information about
the mechanism of instability, and would have been difficult to predict or
identify without the use of the proposed method.
Future work will include
generalizations to 
other controlling sets,
beyond equilibrium points,
and developing control
design strategies to reduce disturbance vulnerability
after identifying the
mode of instability.

\bibliographystyle{ieeetr}
\bibliography{ieee/References}

\begin{thebibliography}{10}

\bibitem{kundur2004definition}
P.~Kundur, J.~Paserba, V.~Ajjarapu, G.~Andersson, A.~Bose, C.~Canizares, N.~Hatziargyriou, D.~Hill, A.~Stankovic, C.~Taylor, T.~Van~Cutsem, and V.~Vittal, ``Definition and classification of power system stability ieee/cigre joint task force on stability terms and definitions,'' {\em IEEE Transactions on Power Systems}, vol.~19, no.~3, pp.~1387--1401, 2004.

\bibitem{chiang1993predicting}
H.-D. Chiang, J.~Tong, and K.~N. Miu, ``Predicting unstable modes in power systems: Theory and computations,'' {\em IEEE Transactions on Power Systems}, vol.~8, no.~4, pp.~1429--1437, 1993.

\bibitem{fouad1991power}
A.~A. Fouad and V.~Vittal, {\em Power System Transient Stability Analysis Using the Transient Energy Function Method}.
\newblock Englewood Cliffs, NJ, USA: Prentice-Hall, 1992.

\bibitem{chiang1988stability}
H.-D. Chiang, M.~W. Hirsch, and F.~F. Wu, ``Stability regions of nonlinear autonomous dynamical systems,'' {\em IEEE Transactions on Automatic Control}, vol.~33, no.~1, pp.~16--27, 1988.

\bibitem{michel1985mechanism}
A.~Michel and V.~Vittal, ``On the mechanism of transient instability of power systems,'' {\em Circuits, Systems and Signal Processing}, vol.~4, no.~3, pp.~413--434, 1985.

\bibitem{ma2023dominant}
R.~Ma, Y.~Zhang, M.~Zhan, K.~Cao, D.~Liu, K.~Jiang, and S.~Cheng, ``Dominant transient equations of grid-following and grid-forming converters by controlling-unstable-equilibrium-point-based participation factor analysis,'' {\em IEEE Transactions on Power Systems}, vol.~39, no.~3, pp.~4818--4834, 2023.

\bibitem{behera1985analytical}
A.~K. Behera, M.~Pai, and P.~Sauer, ``Analytical approaches to determine critical clearing time in multi-machine power systems,'' {\em 24th Conference on Decision and Control (CDC)}, pp.~818--823, 1985.

\bibitem{fouad1984critical}
A.~Fouad, V.~Vittal, and T.~K. Oh, ``Critical energy for direct transient stability assessment of a multimachine power system,'' {\em IEEE Transactions on Power Apparatus and Systems}, vol.~103, no.~8, pp.~2199--2206, 1984.

\bibitem{chiang1988foundations}
H.-D. Chiang, F.~F. Wu, and P.~P. Varaiya, ``Foundations of the potential energy boundary surface method for power system transient stability analysis,'' {\em IEEE Transactions on Circuits and Systems}, vol.~35, no.~6, pp.~712--728, 1988.

\bibitem{treinen1996improved}
R.~T. Treinen, V.~Vittal, and W.~Kliemann, ``An improved technique to determine the controlling unstable equilibrium point in a power system,'' {\em IEEE Transactions on Circuits and Systems I: Fundamental Theory and Applications}, vol.~43, no.~4, pp.~313--323, 1996.

\bibitem{fisher2023stability}
M.~W. Fisher and I.~A. Hiskens, ``Stability of the nonwandering set in the region of attraction boundary under perturbations with application to vulnerability assessment,'' {\em SIAM Journal on Applied Dynamical Systems}, vol.~22, no.~4, pp.~3390--3430, 2023.

\bibitem{fisher2022hausdorff}
M.~W. Fisher and I.~A. Hiskens, ``Hausdorff continuity of region of attraction boundary under parameter variation with application to disturbance recovery,'' {\em SIAM Journal on Applied Dynamical Systems}, vol.~21, no.~1, pp.~327--365, 2022.

\bibitem{fisher2019numerical}
M.~W. Fisher and I.~A. Hiskens, ``Numerical computation of critical system recovery parameter values by trajectory sensitivity maximization,'' {\em 58th Conference on Decision and Control (CDC)}, pp.~8000--8006, 2019.

\bibitem{fisher2025computing}
M.~W. Fisher, ``Computing safety margins of parameterized nonlinear systems for vulnerability assessment via trajectory sensitivities,'' 2025, under review. Preprint available at https://arxiv.org/abs/2501.07498.

\bibitem{hirsch2012differential}
M.~W. Hirsch, {\em Differential Topology}, vol.~33 of \emph{Graduate Texts in Mathematics}.
\newblock Springer-Verlag, 1976.

\bibitem{pugh1983c1}
C.~C. Pugh and C.~Robinson, ``The c1 closing lemma, including hamiltonians,'' {\em Ergodic Theory and Dynamical Systems}, vol.~3, no.~2, pp.~261--313, 1983.

\bibitem{katok1999introduction}
A.~Katok and B.~Hasselblatt, {\em Introduction to the Modern Theory of Dynamical Systems}, vol.~54 of \emph{Encyclopedia of Mathematics and its Applications}.
\newblock Cambridge University Press, 1999.

\bibitem{machowski1997power}
P.~W. Sauer and M.~A. Pai, {\em Power System Dynamics and Stability}.
\newblock 1997.

\bibitem{anderson2003the}
P.~M. Anderson and A.~A. Fouad, {\em The Elementary Mathematical Model}.
\newblock Hoboken, NJ, USA:Wiley-IEEE Press, 2003, pp. 13-52.

\bibitem{ieee2016ieee}
{\em IEEE Std. 421.5-2016, IEEE Recommended Practice for Excitation System Models for Power System Stability Studies}.
\newblock New York: Institute of Electrical and Electronics Engineers, Inc., 2016.

\bibitem{lee2013smooth}
J.~M. Lee, {\em Introduction to Smooth Manifolds}.
\newblock Graduate Texts in Mathematics, Springer, 2 ed., 2013.

\bibitem{palis1969morse}
J.~Palis, ``On morse-smale dynamical systems,'' {\em Topology}, vol.~8, no.~4, pp.~385--405, 1969.

\bibitem{berge1963topological}
C.~Berge, {\em Topological Spaces (Oliver and Boyd, Edinburgh)}.
\newblock 1963.

\bibitem{ortega1990numerical}
J.~M. Ortega, {\em Numerical Analysis: A Second Course}.
\newblock Philadelphia, PA, USA: SIAM, 1990.

\bibitem{sanfelice2014asymptotic}
R.~G. Sanfelice, ``Asymptotic properties of solutions to set dynamical systems,'' {\em 56th Conference on Decision and Control (CDC)}, pp.~2287--2292, 2014.

\end{thebibliography}
\vspace{-1 cm}
\begin{IEEEbiography}[{\includegraphics[width=1in,height=1.25in,clip,keepaspectratio]
     {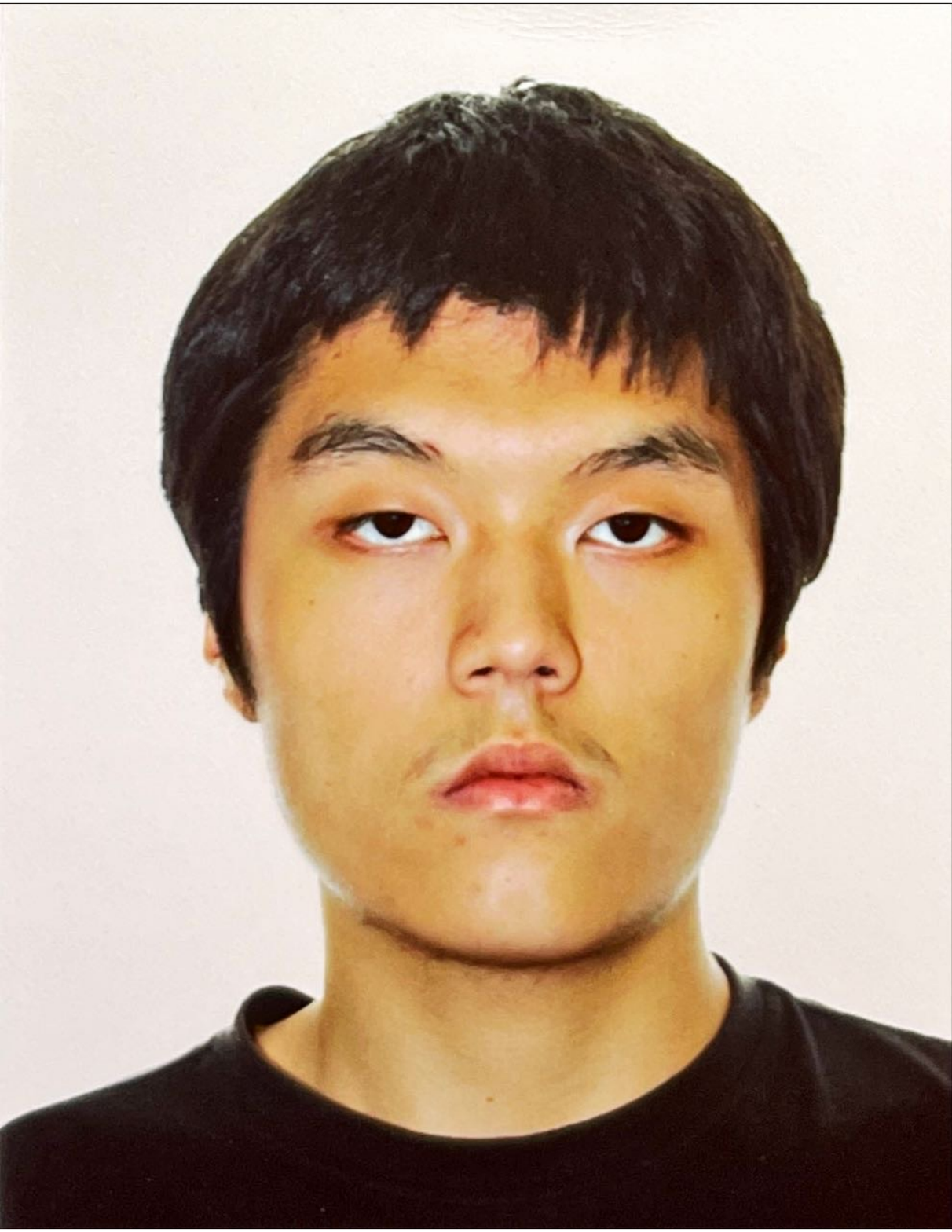}}]{Jinghan Wang}received Bachelor of Mathematics in Honours Applied Mathematics, with Engineering Specialization in Communication and Control, and Pure Mathematics Minor at the University of Waterloo, Ontario, Canada, in 2023. He is currently pursuing Master of Applied Science with Systems and Control Specialization in the Department of Electrical and Computer Engineering at the University of Waterloo, Ontario, Canada. He works in the dynamics, optimization, and control of complex systems group. His research interests include nonlinear stability analysis, control design strategies for improving nonlinear robustness, and power systems.
\end{IEEEbiography}
\vspace{-1 cm}
\begin{IEEEbiography}[{\includegraphics[width=1in,height=1.25in,clip,keepaspectratio]
     {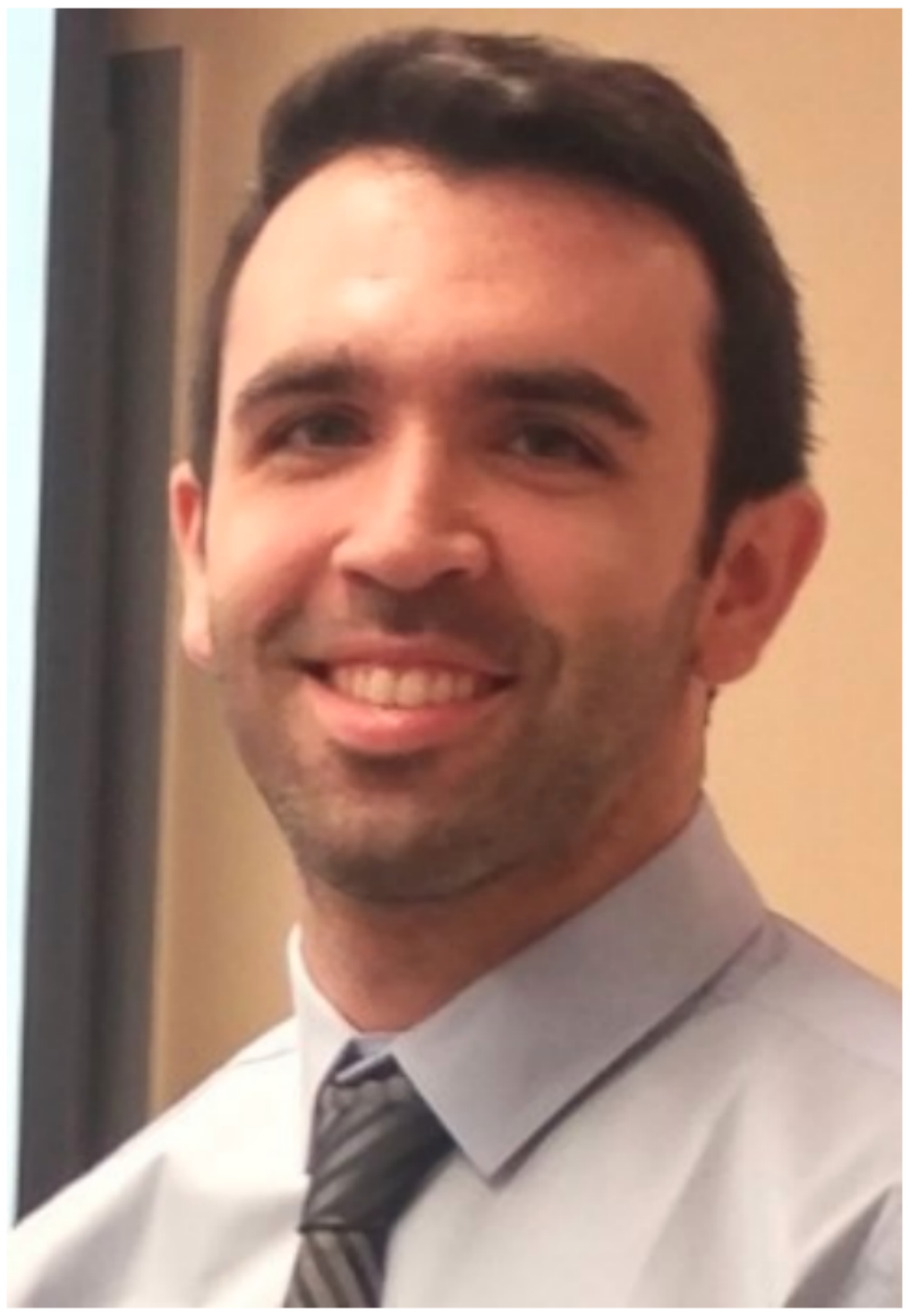}}]{Michael W. Fisher} is an Assistant Professor
  in the Department of Electrical and Computer Engineering at the University of
  Waterloo, Canada.  He was a postdoctoral researcher with
  the Automatic Control and Power System Laboratories at
  ETH Zurich.  He received his Ph.D. in Electrical Engineering:
  Systems at the University of Michigan, Ann Arbor in 2020, and a
  M.Sc. in Mathematics from the same institution in 2017. He received
  his B.A. in Mathematics and Physics from Swarthmore College in 2014.
  His research interests are in dynamics, control, and optimization of
  complex systems.
  He was a finalist for the 2017 Conference on Decision and Control (CDC)
  Best Student Paper Award and a recipient
  of the 2019 CDC Outstanding Student Paper Award.
\end{IEEEbiography}
\end{document}